\documentclass[12pt,a4paper,aps,preprint,superscriptaddress,nofootinbib]{revtex4-1}
\usepackage[utf8]{inputenc}
\usepackage{graphicx}
\usepackage{amssymb}
\usepackage{textcomp}
\usepackage{amsmath}
\usepackage{tabularx}
\usepackage{bm}
\usepackage{times}
\usepackage{color}
\usepackage{slashed}
\usepackage{multirow}
\usepackage{verbatim}
\usepackage{cancel}
\usepackage{subfigure}
\usepackage{ulem}

\usepackage[colorlinks=true, pdfstartview=FitV, linkcolor=blue, citecolor=blue, urlcolor=blue]{hyperref}
\allowdisplaybreaks[4]

\def\lsim{\mathrel{\raise.3ex\hbox{$<$\kern-.75em\lower1ex\hbox{$\sim$}}}}
\def\gsim{\mathrel{\raise.3ex\hbox{$>$\kern-.75em\lower1ex\hbox{$\sim$}}}}

\newcommand{\be}{\begin{equation}}
\newcommand{\ee}{\end{equation}}
\newcommand{\ba}{\begin{array}}
\newcommand{\ea}{\end{array}}
\newcommand{\bea}{\begin{eqnarray}}
\newcommand{\eea}{\end{eqnarray}}

\newcommand{\epem}{e^+e^-}

\definecolor{orange}{rgb}{1,0.5,0}

\usepackage[table]{xcolor}
\usepackage{colortbl}
\definecolor{mygray}{gray}{.95}

\begin{document}

\title{Constraining nonstandard neutrino interactions at electron colliders}

\author{Jiajun Liao}
\email{liaojiajun@mail.sysu.edu.cn}
\affiliation{
School of Physics, Sun Yat-Sen University, Guangzhou 510275, China
}

\author{Yu Zhang}
\email{dayu@ahu.edu.cn}
\affiliation{
	Institutes of Physical Science and Information Technology, Anhui University, Hefei 230601, China
}
\affiliation{
	School of Physics and Materials Science, Anhui University, Hefei 230601,China
}

\begin{abstract}
The presence of nonstandard neutrino interactions (NSI) has a large effect on the precision measurements at next generation neutrino oscillation experiments. Other type of experiments are needed to constrain the NSI parameter space.
We study the constraints on NSI with electrons from current and future $e^+e^-$ collider experiments including Belle II, STCF and CEPC. We find that Belle II and STCF will provide competitive and complementary bounds on electron-type NSI parameters as compared to the current global analysis, and strong improvements in the constraints on tau-type NSI. In addition, CEPC alone will impose stringent constraints on the parameter space of NSI with electrons. We find that the degeneracy between the left-handed (vector) and right-handed (axial-vector) NSI parameters can be lifted by combining the data from three different running modes at CEPC, and the limits on $\epsilon_{ee}^{eL}$ ($\epsilon_{ee}^{eV}$) and $\epsilon_{ee}^{eR}$ ($\epsilon_{ee}^{eA}$) can reach 0.002 at CEPC even if both the left and right-handed NSI parameters are present.
\end{abstract}

\maketitle

\section{Introduction}
\label{sec:Intro}
Neutrino oscillation has been well established after successful measurements from a variety of neutrino experiments using solar, atmospheric, reactor, and accelerator neutrinos in the last two decades, and current data from most neutrino oscillation experiments can be successfully explained in a three neutrino oscillation framework~\cite{Tanabashi:2018oca}. Since the explanation of neutrino oscillations requires the neutrino masses to be nonzero, the observation of neutrino oscillations provides a clear evidence of new physics beyond the Standard Model (SM). Next generation neutrino oscillation experiments,such as DUNE~\cite{Acciarri:2015uup}, Hyper-Kamiokande~\cite{Abe:2018uyc} and JUNO~\cite{An:2015jdp} will not only do a high precision measurement of the oscillation parameters in the standard three neutrino oscillation framework, but also reach the sensitivity to probe new physics in the neutrino sector. A model-independent way of studying new physics in neutrino experiments was given in the effective field theory framework of nonstandard neutrino interactions (NSI);  for reviews see Refs.~\cite{Ohlsson:2012kf, Miranda:2015dra, Farzan:2017xzy}. In this framework, NSI are typically described by dimension-six four-fermion operators of the form~\cite{Wolfenstein:1977ue, Guzzo:1991hi},
\begin{align}
\mathcal{L}_\text{NSI}^\text{NC} &=-2\sqrt{2}G_F
\epsilon^{f C}_{\alpha\beta} 
\left[ \overline{\nu_\alpha} \gamma^{\rho} P_L \nu_\beta \right] 
\left[ \bar{f} \gamma_{\rho} P_C f \right]\,,
\\
\mathcal{L}_\text{NSI}^\text{CC} &=-2\sqrt{2}G_F
\epsilon^{ff^\prime C}_{\alpha\beta} 
\left[ \overline{\nu_\beta} \gamma^{\rho} P_L \ell_\alpha \right] 
\left[ \bar{f^\prime} \gamma_{\rho} P_C f \right]\,,
\label{eq:NSI}
\end{align}
where $\alpha$, $\beta$ label the lepton flavors ($e, \mu, \tau$), $f$ denotes the fermion fields ($u,d,e$), and $C$ indicates the chirality ($L, R$), $\epsilon^{ff^\prime C}_{\alpha\beta}$ are dimensionless parameters that characterize the strength of the new interactions in units of the Fermi constant $G_F$. The neutral current (NC) NSI will affect neutrino propagation in matter, and the charged current (CC) NSI mainly affect neutrino production and detection. Both the NC and CC NSI at neutrino oscillation experiments have been extensively studied in the literature; see the references in Ref.~\cite{Dev:2019anc}. Here we focus on NC NSI since the bounds on the CC NSI are generally much stronger than the NC NSI~\cite{Biggio:2009nt}. The presence of NC NSI have a large impact on the determination of the mass ordering, CP violation and $\theta_{23}$ octant at neutrino oscillation experiments~\cite{GonzalezGarcia:2001mp, Miranda:2004nb, Coloma:2011rq, Friedland:2012tq, Masud:2015xva, deGouvea:2015ndi, Coloma:2015kiu, Liao:2016hsa, Forero:2016cmb, Masud:2016bvp, Ge:2016dlx, Blennow:2016etl, Agarwalla:2016fkh, Fukasawa:2016lew, Deepthi:2016erc, Liao:2016orc}. In particular, due to the presence of the generalized mass ordering degeneracy induced by NSI, the neutrino mass ordering cannot be resolved by neutrino oscillation experiments alone~\cite{Coloma:2016gei}. In order to improve the sensitivity of neutrino oscillation experiments in the presence of NSI, other types of experiments such as neutrino scattering or collider experiments are needed to constrain the NSI parameter space~\cite{Coloma:2017egw}. 

The constraints on NSI with quarks have been well studied in the literature~\cite{Gonzalez-Garcia:2013usa, Esteban:2018ppq}. In particular, the recent measurements of the coherent neutrino nucleus scattering by the COHERENT experiment~\cite{Akimov:2017ade, Akimov:2020pdx} has been used to impose constraints on NSI with the $u$ and $d$ quarks, and the generalized mass ordering degeneracy due to NSI with quarks can be excluded by more than $3\sigma$ with the COHERENT data~\cite{Coloma:2017ncl, Coloma:2019mbs}. In addition, the parameter space of NSI with quarks can be further constrained by using the monojet events at LHC~\cite{Friedland:2011za, Choudhury:2018xsm}. However,
there are models that predict NSI with only electrons but not with quarks in the literature; see e.g., Refs.~\cite{Forero:2016ghr, Babu:2019mfe, Liao:2019qbb}. Due to degeneracies between the left and right-handed NSI parameters, constraints on NSI with electrons are still weak as compared to NSI with quarks. 

The Lagrangian of NC NSI with electrons can be written as, 
\begin{eqnarray}
\label{eq:NSIee}
\mathcal{L}_{\rm{NSI}}^{\rm NC,e}
&=&-2\sqrt{2}G_F\epsilon_{\alpha\beta}^{eL}(\bar{\nu}_\alpha\gamma^\mu P_L \nu_\beta) (\bar{e}\gamma_\mu P_L e)
-2\sqrt{2}G_F\epsilon_{\alpha\beta}^{eR}(\bar{\nu}_\alpha\gamma^\mu P_L \nu_\beta) (\bar{e}\gamma_\mu P_R e)
\\
&=&-\sqrt{2}G_F\epsilon_{\alpha\beta}^{eV}(\bar{\nu}_\alpha\gamma^\mu P_L \nu_\beta) (\bar{e}\gamma_\mu e)
+\sqrt{2}G_F\epsilon_{\alpha\beta}^{eA}(\bar{\nu}_\alpha\gamma^\mu P_L \nu_\beta) (\bar{e}\gamma_\mu \gamma^5 e)\,,
\end{eqnarray}
where    
\begin{equation}
\label{eq:lr2va}
\epsilon_{\alpha\beta}^{eV} \equiv \epsilon_{\alpha\beta}^{eL} + \epsilon_{\alpha\beta}^{eR}\,, 
\quad \epsilon_{\alpha\beta}^{eA} \equiv \epsilon_{\alpha\beta}^{eL} - \epsilon_{\alpha\beta}^{eR}\,,
\end{equation}
with 
$\epsilon^{eV}_{\alpha\beta}$ ($\epsilon^{eA}_{\alpha\beta}$) describing  the
strength of the new vector (axial-vector) interactions between electrons. NSI with electrons will lead to new contributions to the monophoton process $e^+ e^- \to \nu \bar\nu \gamma$ at electron colliders. An early study of NSI with electrons using the monophoton events at LEP has been given in Ref.~\cite{Berezhiani:2001rs}, and constraints on NSI with electrons from neutrino scattering,  solar and reactor neutrino experiments are performed in Refs.~\cite{Davidson:2003ha, Barranco:2005ps, Bolanos:2008km}. A combination of the data from LEP, neutrino scattering, solar and reactor neutrino experiments can be found in Refs.~\cite{Barranco:2007ej, Forero:2011zz}. 
The $e^+e^-$ collider experiment LEP can help in removing degeneracy in the global analysis of constraints on NSI parameters. Moreover, the current constraints on tau-type NSI with electrons are mainly coming from LEP. However, due to cancellation between the left-handed (vector) and right-handed (axial-vector) NSI parameters, the allowed ranges of NSI with electrons are still large if both the left and right-handed NSI parameters exist. Considering future $e^+e^-$ collider experiment will provide much more data than LEP, we study the constraints on NSI with electrons from current $e^+e^-$ collider experiment Belle II~\cite{Abe:2010gxa}, and the proposed future $e^+e^-$ collider experiments including the Super Tau Charm Factory (STCF)~\cite{Luo:2019xqt} and CEPC~\cite{CEPCStudyGroup:2018ghi} 
in this work. 

This paper is organized as follows. In Sec.~\ref{sec:experiments} we describe the signal and backgrounds of probing NSI at electron colliders. In Sec.~\ref{sec:sllmZ} we present the constraints on NSI at Belle II and STCF that are operated with the center-of-mass (CM) energy $\sqrt{s}\ll M_Z$. In Sec.~\ref{sec:sgZ} we consider the constraints on NSI at CEPC that is operated with $\sqrt{s}\geq M_Z$. Our conclusions are drawn in Sec.~\ref{sec:Con}. 

\section{Signals and backgrounds at electron colliders}
\label{sec:experiments}

\begin{figure}[htbp]
	\centerline{\includegraphics[width=15cm]{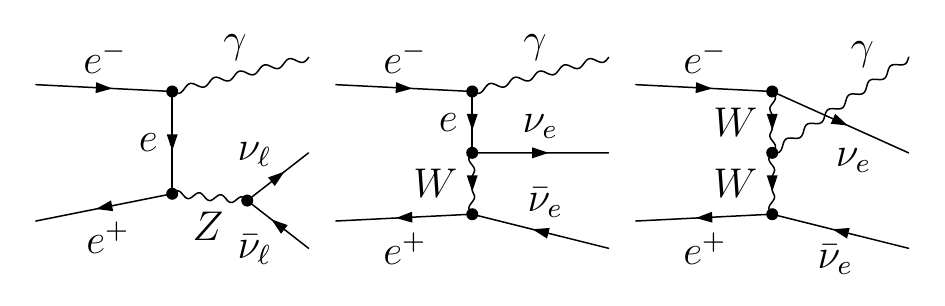}}
	\caption{The tree-level Feynman diagrams for $e^+ e^-\to \nu_\ell \bar\nu_\ell \gamma$ process at electron colliders. 
		The $W$-boson exchange diagrams only contribute to the 
		$e^+e^-\rightarrow \nu_e\bar{\nu_e}\gamma$ process, and 
		the $Z$-boson exchange diagram can contribute to all neutrino flavors with $\ell=e, \mu, \tau$. 
	}
	\label{fig:feynman-irbg}
\end{figure}

The neutrinos can be measured at $e^+e^-$ colliders via
the production of single photon  associated with a neutrino-antineutrino pair, $e^+ e^- \to \nu \bar\nu \gamma$.
In the SM, the tree-level Feynman diagrams for this process are shown in Fig.~\ref{fig:feynman-irbg}.
For  muon  and  tau  neutrinos, only the $Z$-boson exchange diagram contributes; for electron neutrinos, both the $Z$-boson and $W$-boson exchange diagrams contribute.

The cross section for the single photon 
production from $e^+e^-$ annihilation, $e^+ e^- \to \nu \bar\nu \gamma$, can be approximately factorized into the process without
photon emission, $e^+ e^- \to \nu \bar\nu$, times the improved 
Altarelli-Parisi radiator function 
\cite{Nicrosini:1989pn,Montagna:1995wp},
\begin{equation}
\frac{d^2\sigma}{d x_\gamma d z_\gamma}=
H\left(x_\gamma, z_{\gamma}; s\right) \sigma_{0}(s_\gamma)\,,
\label{eq:diffxs}
\end{equation}
where the radiator function 
\begin{equation}
H\left(x_\gamma, z_{\gamma} ; s\right)=\frac{\alpha}{\pi}\frac{1}{x_\gamma} \left[\frac{1+(1-x_\gamma)^2}{1-z_{\gamma}^{2}}- \frac{x_\gamma^2}{2}\right]\,.
\end{equation} 
Here, $E_\gamma$ is the energy of the initial state
radiation (ISR) photon, $\sqrt{s}$ is the CM energy at the colliders, $x_\gamma=2E_\gamma/\sqrt{s}$ is the energy fraction emitted away by ISR, $z_\gamma=\cos\theta_\gamma$ with $\theta_\gamma$ being the polar angle of the photon, $s_\gamma=(1-x_\gamma)s$ is the CM energy of the $\nu\bar\nu$ system, and $\sigma_{0}$ is the cross section of the neutrino pair production without ISR. 

In the SM, $\sigma_{0}$ is given by
\begin{equation}
\begin{aligned}
\sigma_{0}^{\mathrm{SM}}(s)&= \sigma_{W}(s)+\sigma_{Z}(s)+\sigma_{W-Z}(s) \\
&= \frac{G_{F}^{2} s}{12 \pi}\left[2+\frac{2N_{\nu}\left(g_{L}^{2}+g_{R}^{2}\right)}{\left(1-s / M_{Z}^{2}\right)^{2}+\Gamma_{Z}^{2} / M_{Z}^{2}}+\frac{4g_{L}\left(1-s / M_{Z}^{2}\right)}{\left(1-s / M_{Z}^{2}\right)^{2}+\Gamma_{Z}^{2} / M_{Z}^{2}}\right],
\end{aligned}
\label{eq:sigam0}
\end{equation}
where $g_L=-\frac{1}{2}+\sin^2\theta_W$, $g_R=\sin^2\theta_W$, and $N_\nu$ is the number of active neutrinos.
The three  terms in Eq.~(\ref{eq:sigam0}) come from the contribution of the $W$ boson, the $Z$ boson, and 
their interference, respectively. 
The NSI contribution to the cross section $\sigma_{0}$  can be written analogously as 
\begin{equation}
\sigma_{0}^{\mathrm{NSI}}(s) 
= \frac{G_{F}^{2}}{6 \pi} s \left\lbrace  \sum_{\alpha,\beta=e, \mu, \tau} \left[\left((\epsilon_{\alpha\beta}^{eL})^2+(\epsilon_{\alpha\beta}^{eR})^2\right)-2\left(g_{L} \epsilon_{\alpha\beta}^{eL}+g_{R} \epsilon_{\alpha\beta}^{eR}\right) \frac{M_{Z}^{2}\left(s-M_{Z}^{2}\right)}{\left(s-M_{Z}^{2}\right)^{2}+\left(M_{Z} \Gamma_{Z}\right)^{2}}\right] +2\epsilon_{ee}^{eL} \right\rbrace.
\label{eq:sigmaNSI}
\end{equation}

From Eq.~(\ref{eq:sigmaNSI}), we see that there are 12 independent NSI parameters that can be probed at electron colliders. Also, there is no interference between $\epsilon_{\alpha\beta}$ with different flavor indices. Therefore, we consider the NSI parameters with different flavor indices separately. 
In order to constrain the NSI parameter space, we perform a variation of the $\epsilon$ pair with the same flavor indices $(\epsilon_{\alpha\beta}^{eL},\ \epsilon_{\alpha\beta}^{eR})$ (or equivalently $(\epsilon_{\alpha\beta}^{eV},\ \epsilon_{\alpha\beta}^{eA})$), setting other NSI parameters to be zero.
In our analysis, we only consider bounds on the flavor-conserving NSI parameters since the new physics that induces flavor-violating NSI is more likely to be constrained by lepton flavor violating processes\footnote{Note that for the flavor violating NSI parameters, their contributions to the cross section are twice of that from $\epsilon^{eL,R}_{\mu\mu,\tau\tau}$, and the constraints on flavor violating NSI parameters at electron colliders can be easily derived from those on $\epsilon^{eL,R}_{\mu\mu,\tau\tau}$.}.
To constrain the electron-type NSI parameters $\epsilon_{ee}^{eL,R} $,
we perform a variation of the $\epsilon$ pair $(\epsilon_{ee}^{eL},\epsilon_{ee}^{eR})$ at the same time, setting others to be zero, i.e., $\epsilon_{\mu\mu}^{eL,R}=\epsilon_{\tau\tau}^{eL,R}=0$.
Similarly, to constrain the muon-type (tau-type) NSI parameters $\epsilon_{\mu\mu}^{eL,R} $ ($\epsilon_{\tau\tau}^{eL,R} $), we take the electron-type and tau-type (muon-type) NSI parameters to be zero.

When $\sqrt{s}$ is far below the $Z$ resonance ($\sqrt{s} \ll M_Z$), the differential production cross section for the $e^+ e^- \to \nu \bar\nu \gamma$ in the SM can be written as
%
\be
{d\sigma^{\rm SM} \over d x_\gamma d z_\gamma}
=   H(x_\gamma,z_\gamma,s){G_F^2    s_\gamma \over 2\pi    }	C^{\rm SM}\,,
\label{eq:irbg}
\ee
where 
$C^{\rm SM}\equiv g_L^2+g_R^2+\frac{2}{3}g_L+\frac{1}{3}$.
Here we have integrated over the momenta of the
final state neutrinos and summed all three
neutrino flavors.
The contribution from NSI can be expressed as

\be
{d\sigma^{\rm NSI} \over d x_\gamma d z_\gamma}
=  H(x_\gamma,z_\gamma,s){G_F^2    s_\gamma \over 2\pi    }	C^{\rm NSI}\,,
\label{eq:NSI}
\ee
where $C^{\rm NSI}\equiv\frac{1}{3}\sum\limits_{\alpha,\beta=e,\mu,\tau}\Big[(\epsilon_{\alpha\beta}^{eL})^2+(\epsilon_{\alpha\beta}^{eR})^2
+2(g_L \epsilon_{\alpha\beta}^{eL}+g_R \epsilon_{\alpha\beta}^{eR}) \Big]+\frac{2}{3}{\epsilon_{ee}^{eL}}$.
From Eqs.~(\ref{eq:irbg}) and (\ref{eq:NSI}), we see that for $e^+ e^- \to \nu \bar\nu \gamma$ production
with $\sqrt{s}\ll M_Z$, 
\be
{\sigma^{\rm NSI} \over \sigma^{\rm SM}} 
={C^{\rm NSI}\over C^{\rm SM}}\,,
\label{eq:delta}
\ee
The total cross section of the $e^+ e^- \to \nu \bar\nu \gamma$ process is $\sigma=\sigma^{\rm SM}+\sigma^{\rm NSI}$.
Now the iso-$\sigma$ contour is a circle in the $(\epsilon_{\alpha\alpha}^{eL},\ \epsilon_{\alpha\alpha}^{eR})$ plane with the center located at 
\be
(\epsilon_{\alpha\alpha}^{eL},\ \epsilon_{\alpha\alpha}^{eR})=(-g_L-\delta_{\alpha e},-g_R)\,, 
\label{eq:center-LR-low}
\ee
where $\delta_{\alpha e}=1$ (0) for $\alpha=e$ ($\alpha\neq e$).
Similarly, the center of the iso-$\sigma$ circle in the $(\epsilon_{\alpha\alpha}^{eV},\ \epsilon_{\alpha\alpha}^{eA})$ plane is located at
\be
(\epsilon_{\alpha\alpha}^{eV},\ \epsilon_{\alpha\alpha}^{eA})
= (-g_V-\delta_{\alpha e},-g_A-\delta_{\alpha e})\,, 
\label{eq:center-VA-low}
\ee
where $g_V=g_L+g_R=-\frac{1}{2}+2\sin^2\theta_W$, $g_A=g_L-g_R=-\frac{1}{2}$.

For energies above the $Z$ resonance ($\sqrt{s}\geq M_Z$), finite distance effects on the $W$ propagator need to be
considered. These effects can be taken into account by the following substitution \cite{Berezhiani:2001rs,Hirsch:2002uv}:
\be
\begin{aligned}
	\sigma_{W}(s) & \rightarrow \sigma_{W}(s) F_{W}\left(s / M_{W}^{2}\right)\,, \\
	\sigma_{W-Z}(s) & \rightarrow \sigma_{W-Z}(s) F_{W-Z}\left(s / M_{W}^{2}\right)\,,
\end{aligned}
\ee
where
\be
\begin{aligned}
	F_{W}(z) &=\frac{3}{z^{3}}[-2(z+1) \log (z+1)+z(z+2)]\,, \\
	F_{W-Z}(z) &=\frac{3}{z^{3}}\left[(z+1)^{2} \log (z+1)-z\left(\frac{3}{2} z+1\right)\right].
\end{aligned}
\ee

Using the SM couplings
and the contact NSI interaction (\ref{eq:NSIee}), one can get~\cite{Berezhiani:2001rs}
\begin{equation}
\begin{aligned}
\sigma_{0}^{\mathrm{SM}}(s) 
&= \frac{N_{\nu} G_{F}^{2}}{6 \pi} M_{Z}^{4}\left(g_{L}^{2}+g_{R}^{2}\right) \frac{s}{\left[\left(s-M_{Z}^{2}\right)^{2}+\left(M_{Z} \Gamma_{Z}\right)^{2}\right]} \\
&+\frac{G_{F}^{2}}{\pi} M_{W}^{2}\left\{\frac{s+2 M_{W}^{2}}{2 s}\right.
-\frac{M_{W}^{2}}{s}\left(\frac{s+M_{W}^{2}}{s}\right) \log \left(\frac{s+M_{W}^{2}}{M_{W}^{2}}\right) \\
&-g_{L} \frac{M_{Z}^{2}\left(s-M_{Z}^{2}\right)}{\left(s-M_{Z}^{2}\right)^{2}+\left(M_{Z} \Gamma_{Z}\right)^{2}} 
\left.\times\left[\frac{\left(s+M_{W}^{2}\right)^{2}}{s^{2}} \log \left(\frac{s+M_{W}^{2}}{M_{W}^{2}}\right)-\frac{M_{W}^{2}}{s}-\frac{3}{2}\right]\right\}\,,
\end{aligned}
\label{eq:xs0sm}
\end{equation}

\begin{equation}
\begin{aligned}
\sigma_{0}^{\mathrm{NSI}}(s) & 
= \sum_{\alpha,\beta=e, \mu, \tau} \frac{G_{F}^{2}}{6 \pi} s\left[\left((\epsilon_{\alpha\beta}^{eL})^2+(\epsilon_{\alpha\beta}^{eR})^2\right)-2\left(g_{L} \epsilon_{\alpha\beta}^{eL}+g_{R} \epsilon_{\alpha\beta}^{eR}\right) \frac{M_{Z}^{2}\left(s-M_{Z}^{2}\right)}{\left(s-M_{Z}^{2}\right)^{2}+\left(M_{Z} \Gamma_{Z}\right)^{2}}\right] \\
&+\frac{G_{F}^{2}}{\pi} \epsilon_{ee}^{eL} M_{W}^{2}\left[\frac{\left(s+M_{W}^{2}\right)^{2}}{s^{2}} \log \left(\frac{s+M_{W}^{2}}{M_{W}^{2}}\right)-\frac{M_{W}^{2}}{s}-\frac{3}{2}\right]\,.
\end{aligned}
\label{eq:xs0nsi}
\end{equation}
According to the coefficient of the NSI parameter, the Eq.(\ref{eq:xs0nsi})  can also be written as
\begin{equation}
\begin{aligned}
\sigma_{0}^{\mathrm{NSI}}(s) & 
= \left[\sigma_{0}^{\mathrm{NSI}}(s)\right]_1\sum_{\alpha,\beta=e, \mu, \tau}\left((\epsilon_{\alpha\beta}^{eL})^2+(\epsilon_{\alpha\beta}^{eR})^2\right) \\
&+ \left[\sigma_{0}^{\mathrm{NSI}}(s)\right]_2\sum_{\alpha,\beta=e, \mu, \tau}\epsilon_{\alpha\beta}^{eL}
+ \left[\sigma_{0}^{\mathrm{NSI}}(s)\right]_2\frac{g_R}{g_L}\sum_{\alpha,\beta=e, \mu, \tau}\epsilon_{\alpha\beta}^{eR} \\
&+ \left[\sigma_{0}^{\mathrm{NSI}}(s)\right]_3\epsilon_{ee}^{eL}.
\end{aligned}
\label{eq:xs0nsi2}
\end{equation}

Integrating out  the final photon, we can get the total cross section 
for the process $e^+ e^- \to \nu \bar\nu \gamma$ from NSI at the electron colliders with $\sqrt{s}\geq M_Z$
as 
\begin{equation}
\sigma^{\mathrm{NSI}}\left(\epsilon_{e e}^{e L}, \epsilon_{e e}^{e R}\right)=I_{1}\left(\left(\epsilon_{e e}^{e L}\right)^{2}+\left(\epsilon_{e e}^{e R}\right)^{2}\right)+(I_{2}+I_{3}) \epsilon_{e e}^{e L}+I_{2} \frac{g_R}{g_L}\epsilon_{e e}^{e R}
\end{equation}
for electron-type NSI parameter, and 
\begin{equation}
\sigma^{\mathrm{NSI}}\left(\epsilon_{\mu\mu/\tau\tau}^{e L}, \epsilon_{\mu\mu/\tau\tau}^{e R}\right)=I_{1}\left(\left(\epsilon_{\mu\mu/\tau\tau}^{e L}\right)^{2}+\left(\epsilon_{\mu\mu/\tau\tau}^{e R}\right)^{2}\right)+I_{2} \epsilon_{\mu\mu/\tau\tau}^{e L}+I_{2} \frac{g_R}{g_L}\epsilon_{\mu\mu/\tau\tau}^{e R}
\end{equation}
for muon-type or tau-type NSI parameter, 
with the definition of 
\be
I_i\equiv \int dx \int dz_\gamma H(x_\gamma,z_\gamma,s) \left[\sigma_{0}^{\mathrm{NSI}}((s_\gamma)\right]_i.
\ee
The iso-$\sigma$ contour is also a circle  in the NSI parameter plane, of which the center locates at 
\be 
(\epsilon_{ee}^{eL},\ \epsilon_{ee}^{eR})=(-{I_2+I_3 \over 2I_1},-{I_{2} g_R\over 2I_1 g_L}),
\label{eq:center-e-high}
\ee 
or 
\be
(\epsilon_{\mu\mu/\tau\tau}^{e L}, \epsilon_{\mu\mu/\tau\tau}^{e R})=(-{I_2 \over 2I_1},-{I_{2} g_R\over 2I_1 g_L}).
\label{eq:center-mu-high}
\ee
We can see that the coordinates of the circle center is a function of $\sqrt{s}$.

For NSI searches with the monophoton signature at electron colliders,
the background can be classified into two categories:
the irreducible background and the reducible background. 
The irreducible background has the final state containing one photon 
and two neutrinos arising from the SM, which has been discussed above. 
The reducible background comes from a photon produced in 
the final state together with several other
visible particles which are however not detected due to limitations of the detector acceptance. 
Since the reducible background strongly depends on the detector performance, 
we discuss it later in details for each experiment.
%
%
To evaluate the sensitivity to NSI at electron collider experiments, we define a $\chi^2$ as the function of the two NSI parameters, i.e., 
\be
\label{eq:chi2}
\chi^{2}(\epsilon_{\alpha\beta}^{eL},\ \epsilon_{\alpha\beta}^{eR}) \equiv S^2(\epsilon_{\alpha\beta}^{eL},\ \epsilon_{\alpha\beta}^{eR})/(B_{\rm ir} +B_{\rm re}),
\ee 
where $S$, $B_{\rm ir}$ and $B_{\rm re}$ are the number of events in the signal, irreducible and reducible background, respectively.



\section{$e^+e^-$ colliders operated with $\sqrt{s}\ll M_Z$}
\label{sec:sllmZ}
We first consider constraints on NSI with electrons at $\epem$ colliders that are operated with the CM energies
$\sqrt{s}\ll M_Z$, such as Belle II and STCF. Here we present the detailed analyis of these two experiments below.

\subsection{Belle \uppercase\expandafter{\romannumeral2}}\label{sec:belle2}

At Belle II, photons and electrons can be detected in the 
Electromagnetic Calorimeter (ECL), which covers a polar angle region of $(12.4-155.1)^{\circ}$
and has inefficient gaps between the endcaps and the barrel for polar angles
between $(31.4-32.2)^{\circ}$ and $(128.7-130.7)^{\circ}$ in the lab frame  \cite{Kou:2018nap}.
For monophoton signature at Belle II, the reducible background comes from 
two major parts:
one is mainly due to the lack of polar angle  coverage 
of the ECL near the beam directions, which is referred to 
as the ``bBG''; 
the other one is {mainly} because of the gaps between 
the three segments in the ECL detector,  
which is referred to as the ``gBG''. 
The bBG  arises from the electromagnetic processes $e^+e^-\to \gamma +\slashed{X}$,
mainly containing $e^+e^-\to\gamma\slashed{\gamma}(\slashed{\gamma})$ and $e^+e^-\to\gamma\slashed{e}^+\slashed{e}^-$,
in which  except the detected photon all the other final state particles 
are emitted along the beam directions with 
$\theta>155.1^{\circ}$ or $\theta<12.4^{\circ}$ in the lab frame. 
At Belle II, we adopt the detector cuts for the final detected photon 
(hereafter the ``{\it basic cuts}"):
$12.4^{\circ}<\theta_\gamma<155.1^{\circ}$ in the lab frame.

For the Belle II detector, which is asymmetric, 
the maximum energy of the monophoton events 
in the bBG in the CM frame,
$E_\gamma^m$,  is given by \cite{Liang:2019zkb, Zhang:2020fiu}
(if not exceeding $\sqrt{s}/2$) 
\be
E_\gamma^m(\theta_\gamma) = 
\frac{ \sqrt{s}(A\cos\theta_1-\sin\theta_1)}
{A(\cos\theta_1-\cos\theta_\gamma)-(\sin\theta_\gamma+\sin\theta_1)},
\label{eq:bBG}
\ee 
where  
all angles are given in the CM frame, 
and $A=(\sin\theta_1-\sin\theta_2)/(\cos\theta_1-\cos\theta_2)$, 
with $\theta_1$ and $\theta_2$ being 
the polar angles corresponding to 
the edges of the ECL detector.
In order to remove the above bBG, we use the detector 
cut 
\be
E_\gamma > E_\gamma^m  
\ee
for the final monophoton (hereafter the {\it``bBG cut"}).

Since the gaps in the  ECL are significantly away from the beam direction,
in the gBG, the monophoton energy can be quite large in the central 
$\theta_{\gamma}$ region.
The gBG simulations have been presented by 
Ref.\ \cite{Kou:2018nap} in searching for an  
invisibly decaying dark photon.
In order to optimize the detection efficiency for different masses of
the dark photon, they design two different sets of detector cuts:
the {\it``low-mass cut"} and {\it``high-mass cut"}.
By integrating the differential cross section in Eq.~(\ref{eq:irbg})
in the phase space regions under these two different detector cuts,
and assuming photon detection efficiency as 95\% \cite{Kou:2018nap},
with 50 ab$^{-1}$ integrated luminosity about 2280 (15230) irreducible BG 
events with the {\it``low-mass cut"} ({\it``high-mass cut"})  
can be reached at Belle II\cite{Liang:2019zkb}. 
For the reducible background, 
it is found that about 300 (25000) gBG events 
survived the  {\it``low-mass cut"} ({\it``high-mass cut"}) with 
20 fb$^{-1}$ integrated luminosity  \cite{Kou:2018nap}.

\begin{figure}[htbp]
	\begin{centering}
		\includegraphics[width=0.45\columnwidth]{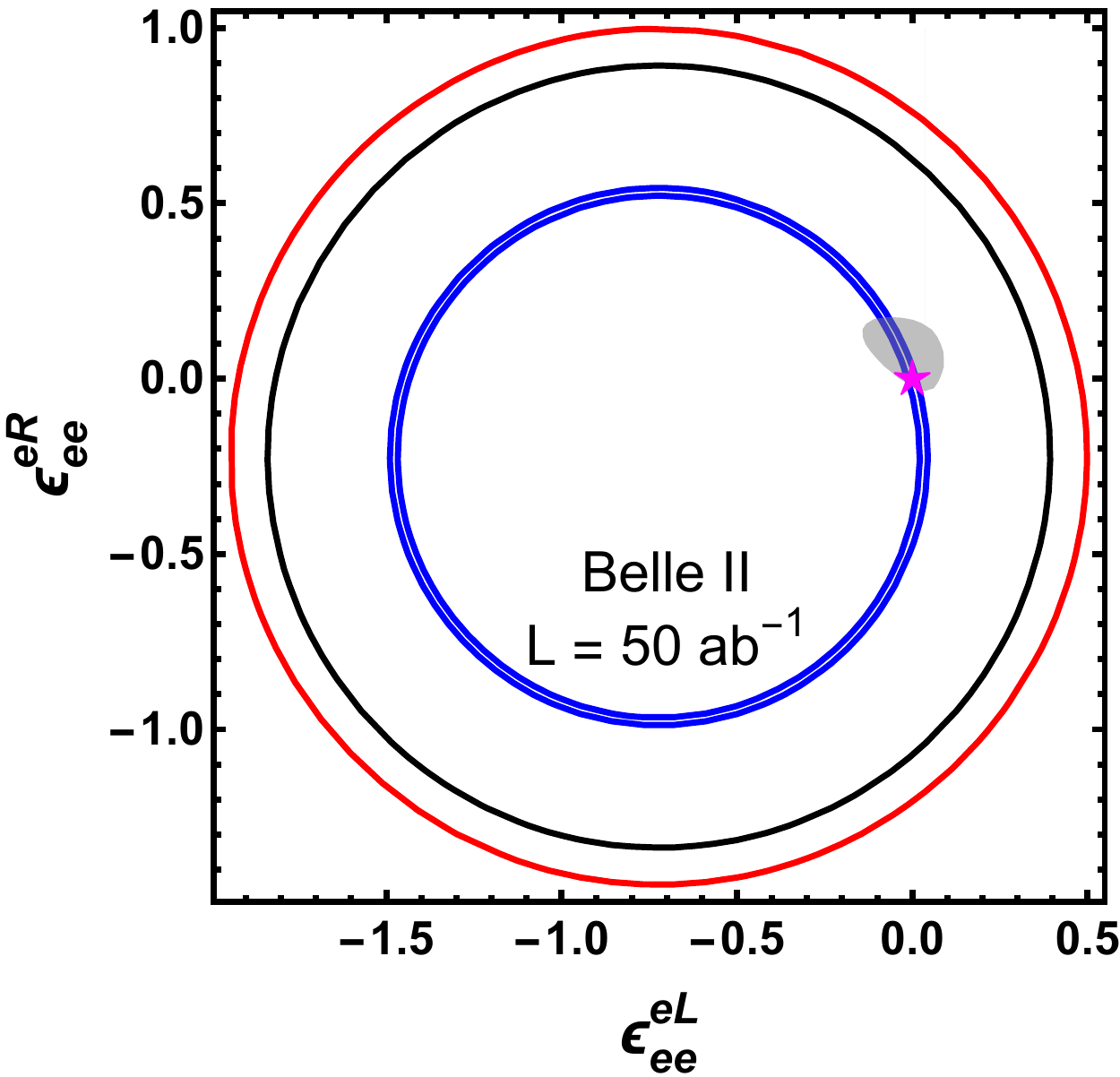}
		\includegraphics[width=0.45\columnwidth]{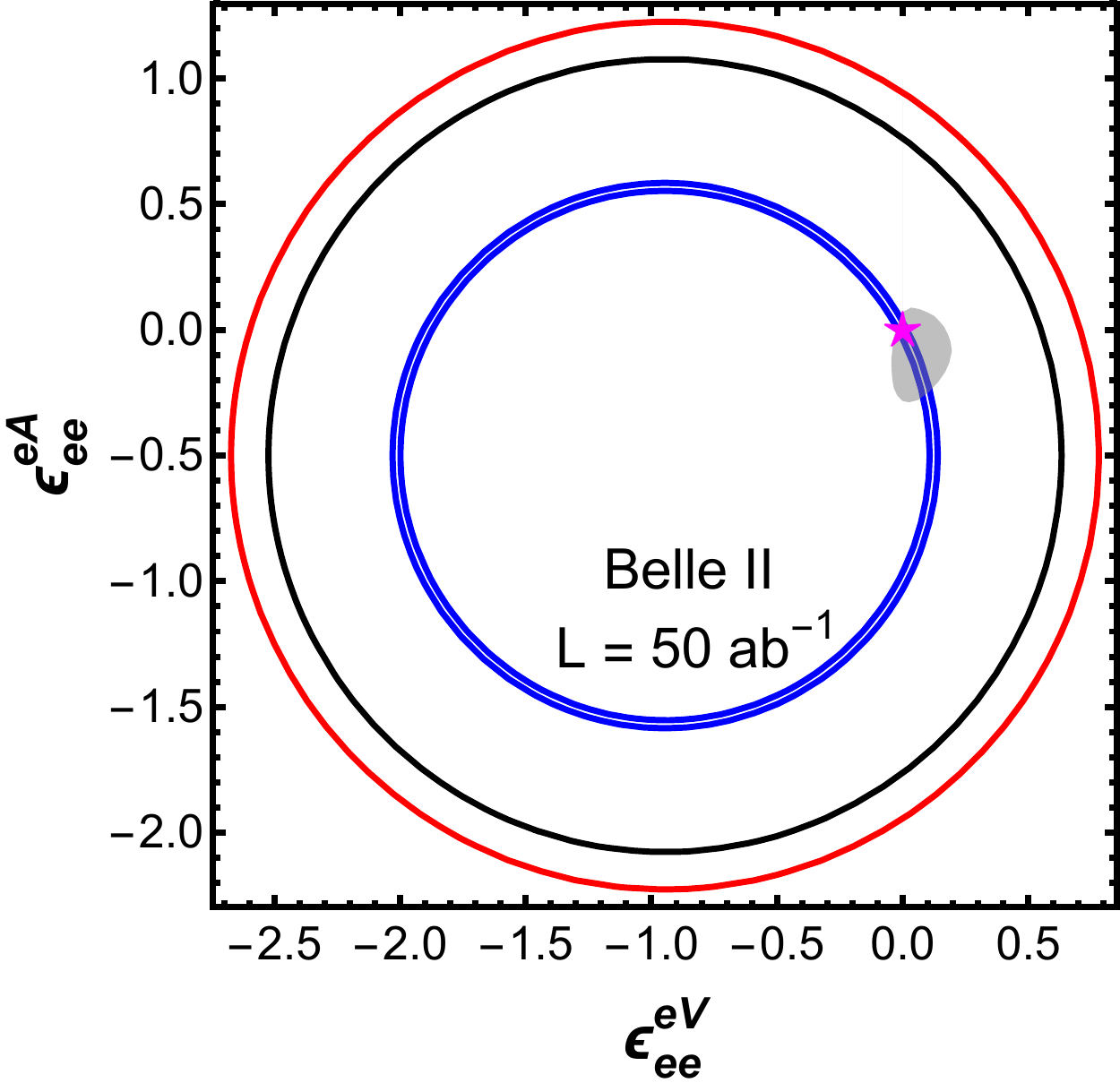}
		\caption{The allowed  90\% C.L. region for electron-type neutrino NSI in the planes of $(\epsilon_{ee}^{eL},\ \epsilon_{ee}^{eR})$ (left panel) and $(\epsilon_{ee}^{eV},\ \epsilon_{ee}^{eA})$ (right panel) at  L = 50 ab$^{-1}$ Belle II  with {\it``low-mass cut"} (black lines), {\it``high-mass cut"} (red lines) and ``{\it bBG cuts}" (blue lines), respectively.
			The  allowed 90\% C.L. regions  arising	from the global analysis of the LEP, CHARM II, LSND, and reactor data~\cite{Barranco:2007ej},  are shown in the shaded gray regions.}
		\label{fig:belle2ee}
	\end{centering}
\end{figure}

\begin{figure}[htbp]
	\begin{centering}
		\includegraphics[width=0.45\columnwidth]{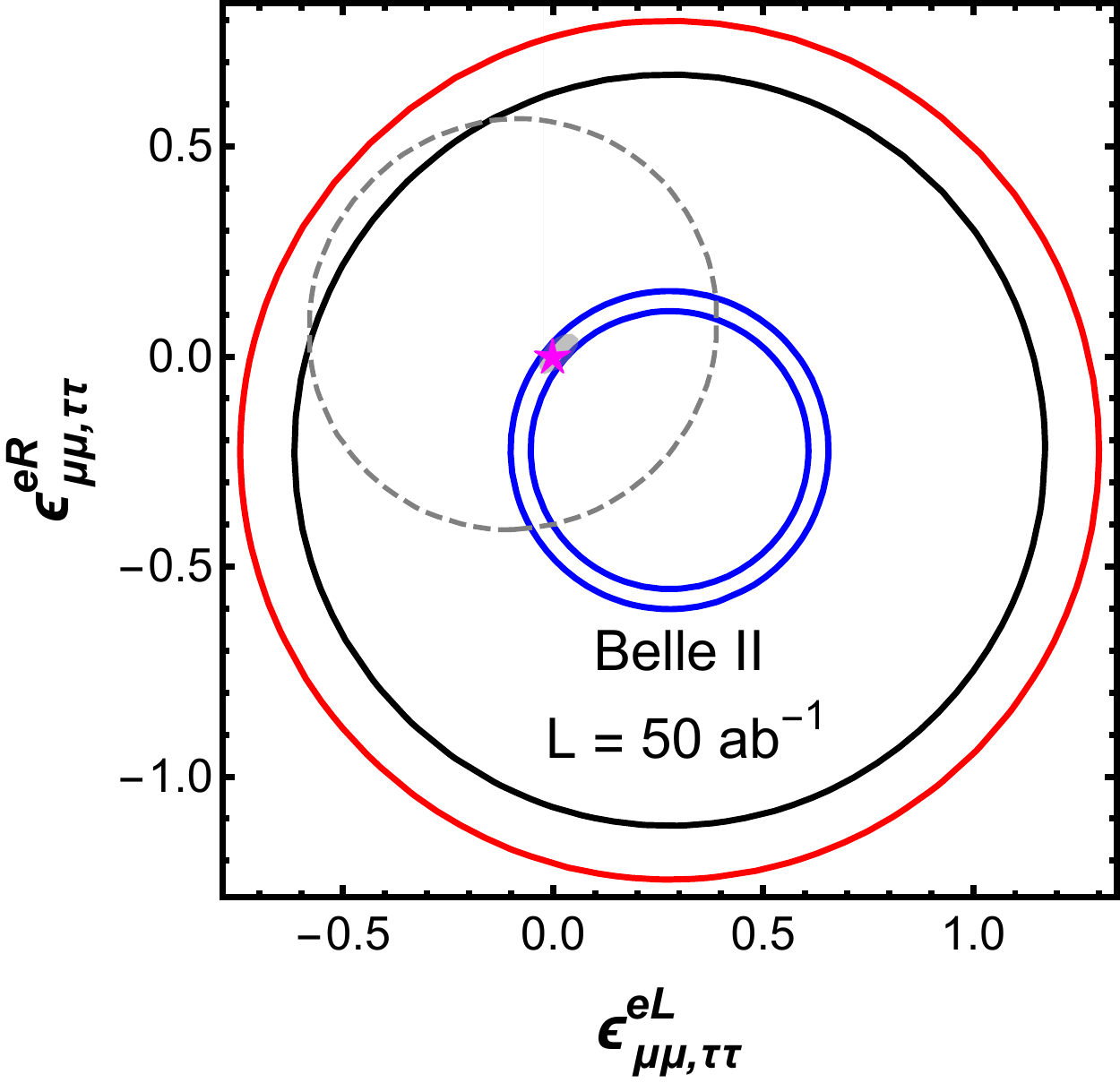}
		\includegraphics[width=0.45\columnwidth]{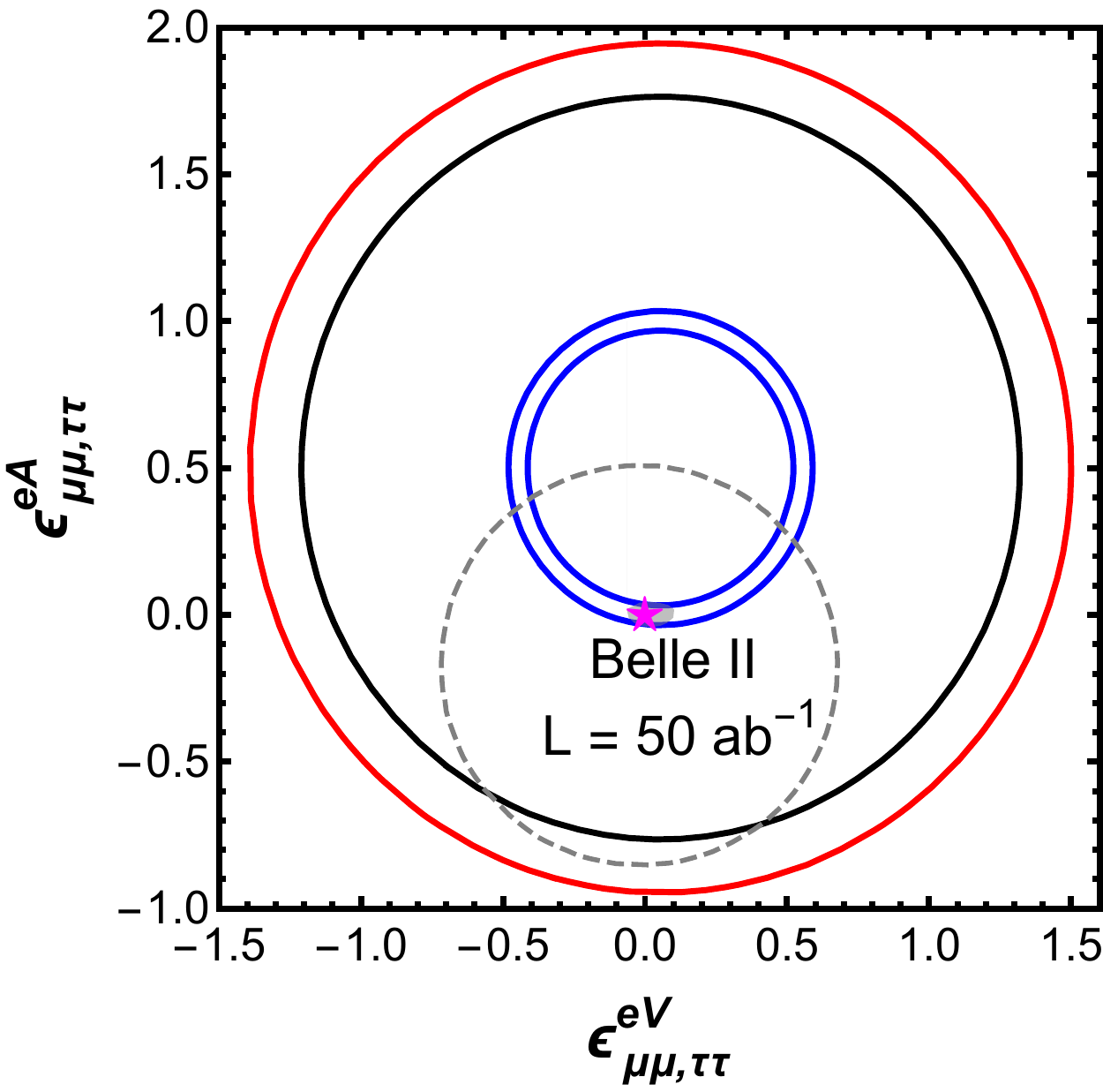}
		\caption{The allowed  90\% C.L. region for muon-type and tau-type neutrino NSI in the planes of $(\epsilon_{\mu\mu,\tau\tau}^{eL},\ \epsilon_{\mu\mu,\tau\tau}^{eR})$ (left panel) and $(\epsilon_{\mu\mu,\tau\tau}^{eV},\ \epsilon_{\mu\mu,\tau\tau}^{eA})$ (right panel) at L = 50 ab$^{-1}$ Belle II  with ``low-mass" cut (black lines), ``high-mass" cut (red lines) and ``{\it bBG cuts}" (blue lines), respectively.
			The  allowed 90\% C.L. regions for muon-type arising
			from the global analysis of the LEP, CHARM II, LSND, and reactor data \cite{Barranco:2007ej},  are shown in the shaded gray regions, and the allowed 95\% C.L. regions for tau-type NSI arising from the LEP data are shown in dashed gray lines.}
		\label{fig:belle2mm}
	\end{centering}
\end{figure}

With the  {\it``low-mass cut"} ({\it``high-mass cut"}) at 50 ab$^{-1}$ Belle II, the numbers of event from irreducible and 
reducible background are $B_{\rm ir}=2280\ (15230)$ and $B_{\rm re}=7.5 \times 10^5\ (6.25\times 10^7)$.
From Eqs.~(\ref{eq:delta}) and (\ref{eq:chi2}), we can get $\chi^2 = \delta^2 B^2_{\rm ir}/(B_{\rm ir}+B_{\rm re})$, where $\delta\equiv {C^{\rm NSI}/ C^{\rm SM}}$.
By solving $\chi^{2} (\epsilon_{\alpha\alpha}^{eL/V},\ \epsilon_{\alpha\alpha}^{eR/A})-\chi^{2} (0,0)=4.61$ for each specific pair,
we present 90\% C.L. allowed  regions at Belle II with the {\it``low-mass cut"}  and {\it``high-mass cut"} 
in Fig. \ref{fig:belle2ee}  for electron-type and Fig. \ref{fig:belle2mm} for muon/tau-type NSI parameters, respectively.
Notice that, the constraints on muon-type NSI parameters from electron colliders are the same as tau-type NSI parameters.
The allowed region under the {\it``low-mass cut"} lies between two concentric black circles, and 
under the {\it``high-mass cut"} in the red circle.
We can see that the centers of the circles are all
located at the $(\epsilon_{\alpha\alpha}^{eL},\ \epsilon_{\alpha\alpha}^{eR})
= (-g_L-\delta_{\alpha e},-g_R)$ and $(\epsilon_{\alpha\alpha}^{eV},\ \epsilon_{\alpha\alpha}^{eA})
= (-g_V-\delta_{\alpha e},-g_A-\delta_{\alpha e})$,
which agree with Eqs. (\ref{eq:center-LR-low})
and (\ref{eq:center-VA-low}). 

We also show together the $(\epsilon_{ee/\mu\mu}^{eL},\ \epsilon_{ee/\mu\mu}^{eR})$ allowed regions at 90\% C.L. from the global analysis with data of the LEP, CHARM II, LSND, and reactor neutrino experiments \cite{Barranco:2007ej}, which are shown in the shaded gray regions.
The constraints on the parameters $(\epsilon_{\tau\tau}^{eL},\ \epsilon_{\tau\tau}^{eR})$ 
only come from the LEP data \cite{Barranco:2007ej}, and are shown in dashed gray lines.
The allowed regions from the global analysis for $(\epsilon_{\alpha\alpha}^{eV},\ \epsilon_{\alpha\alpha}^{eA})$  are translated from the results of $(\epsilon_{\alpha\alpha}^{eL},\ \epsilon_{\alpha\alpha}^{eR})$ in Ref. \cite{Barranco:2007ej} by using the relations in Eq.~(\ref{eq:lr2va}).
We can see that the sensitivity under the {\it``low-mass cut"} is better than the one under the {\it``high-mass cut"}.
Using either {\it``low-mass cut"} or {\it``high-mass cut"}, the 50 ab$^{-1}$ Belle II will be not competitive with current global analysis 
$(\epsilon_{ee, \mu\mu}^{eL},\ \epsilon_{ee, \mu\mu}^{eR})$ and $(\epsilon_{ee, \mu\mu, \tau\tau}^{eV},\ \epsilon_{ee, \mu\mu, \tau\tau}^{eA})$, but can be  complementary with existing LEP data for $(\epsilon_{\tau\tau}^{eL},\ \epsilon_{\tau\tau}^{eR})$.

In order to compare with other experiments 
where detailed simulations with gBG 
are not available, we present the limits with 
gBG ignored for illustration.
We use the ``{\it bBG cut}" defined above to remove the reducible 
background events.
At this time, if gBG is not taken into account, the remain background 
events survived the ``{\it bBG cut}" come from  irreducible backgrounds.
One can get $\chi^2 = \delta^2 \sigma^{SM}L$, where $\sigma^{SM}$ is the 
total cross section by integrating Eq.~(\ref{eq:irbg}) with the ``{\it bBG cut}".
The 90\% C.L. allowed regions analyzed with the ``{\it bBG cut}"  
are shown in Figs. \ref{fig:belle2ee} and \ref{fig:belle2mm}, which lie between 
the two concentric blue circles. One can find that the constraints is obvious stronger 
than the one when gBG is considered under the {\it``low-mass cut"}.
Under ``{\it bBG cut}", the 50 ab$^{-1}$ Belle II with gBG ignored can provide 
a good complement with current global to constrain the NSI parameter.

\subsection{STCF}

The proposed STCF~\cite{Luo:2019xqt} in China is a symmetric double ring electron-positron collider. It is the next generation tau charm facility and  successor  of  the BESIII  experiment,
and designed to have CM energy ranging from 2 to 7 GeV.
Due to the similarity of STCF and BESIII, we present a preliminary projection limit for STCF, 
assuming the same detector performance with BESIII.
For the final photon, we adopt the cuts:
$E_\gamma > 25 $\ MeV in the barrel ($|z_\gamma|<0.8$)  or 
$E_\gamma > 50 $\ MeV in the end-caps ($0.86<|z_\gamma|<0.92$),
following the cuts used by the BESIII Collaboration \cite{Ablikim:2017ixv},
which are defined as the ``{\it basic cuts}" hereafter.

For STCF, which are symmetric, 
the maximum energy of the monophoton events in the
bBG in the CM frame, 
$E_\gamma^m$,  is given by \cite{Liu:2019ogn}
\be
E_\gamma^m(\theta_\gamma) = 
\sqrt{s}\left(\frac{\sin\theta_\gamma}{\sin\theta_b}\right)^{-1},
\label{eq:bBG-besiii}
\ee 
where $\theta_b$ is the polar angle corresponding to the edge
of the detector. 
We will collectively refer to the ``{\it basic cuts}" and cut $E_\gamma > E_\gamma^m$ as the ``{\it advanced cuts}" hereafter.
After considering all the boundary of the subdetectors, we take $\cos\theta_b=0.95$
at the STCF \cite{Liu:2018jdi}.

\begin{figure}[htbp]
	\begin{centering}
		\includegraphics[width=0.45\columnwidth]{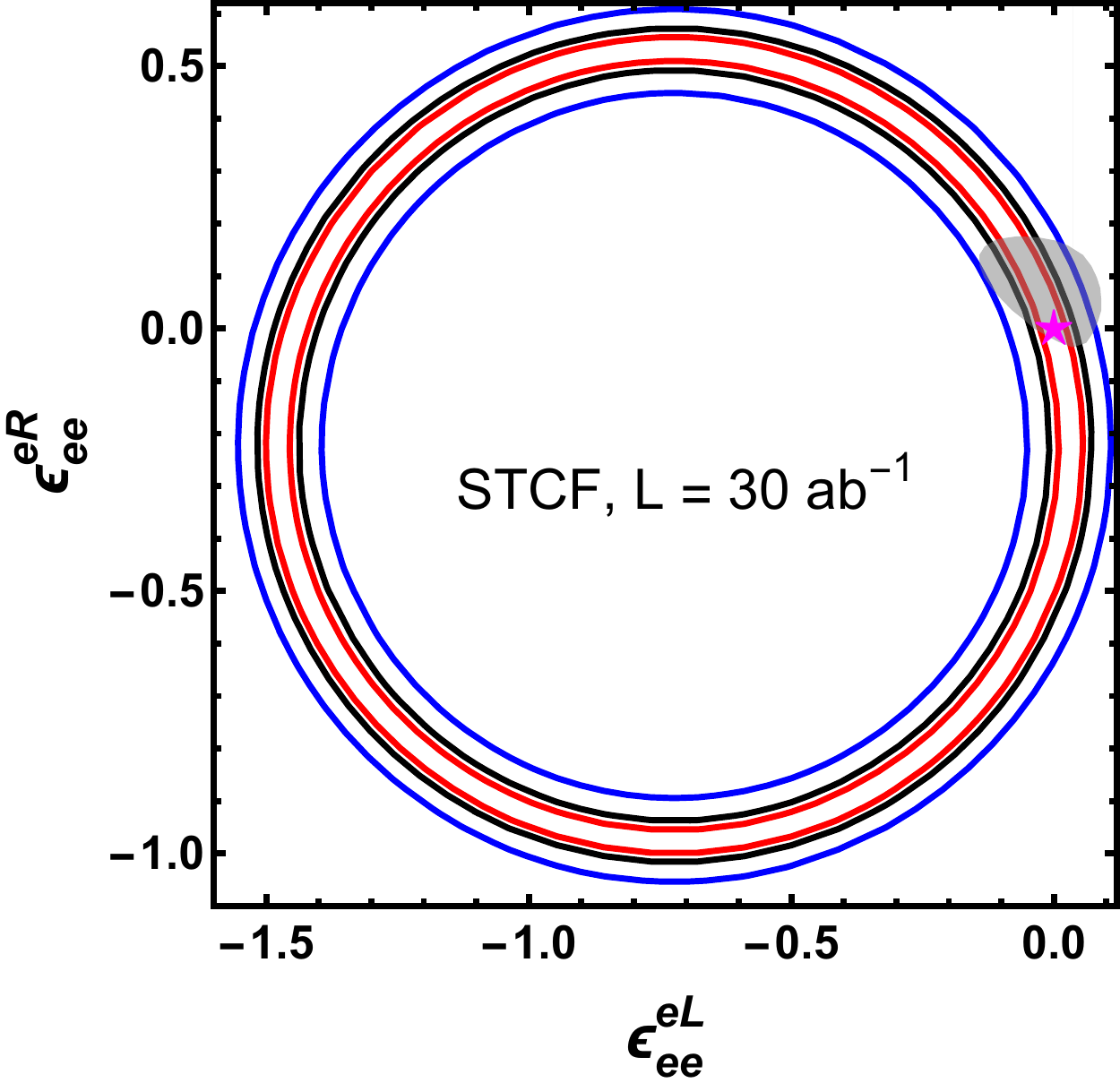}
		\includegraphics[width=0.45\columnwidth]{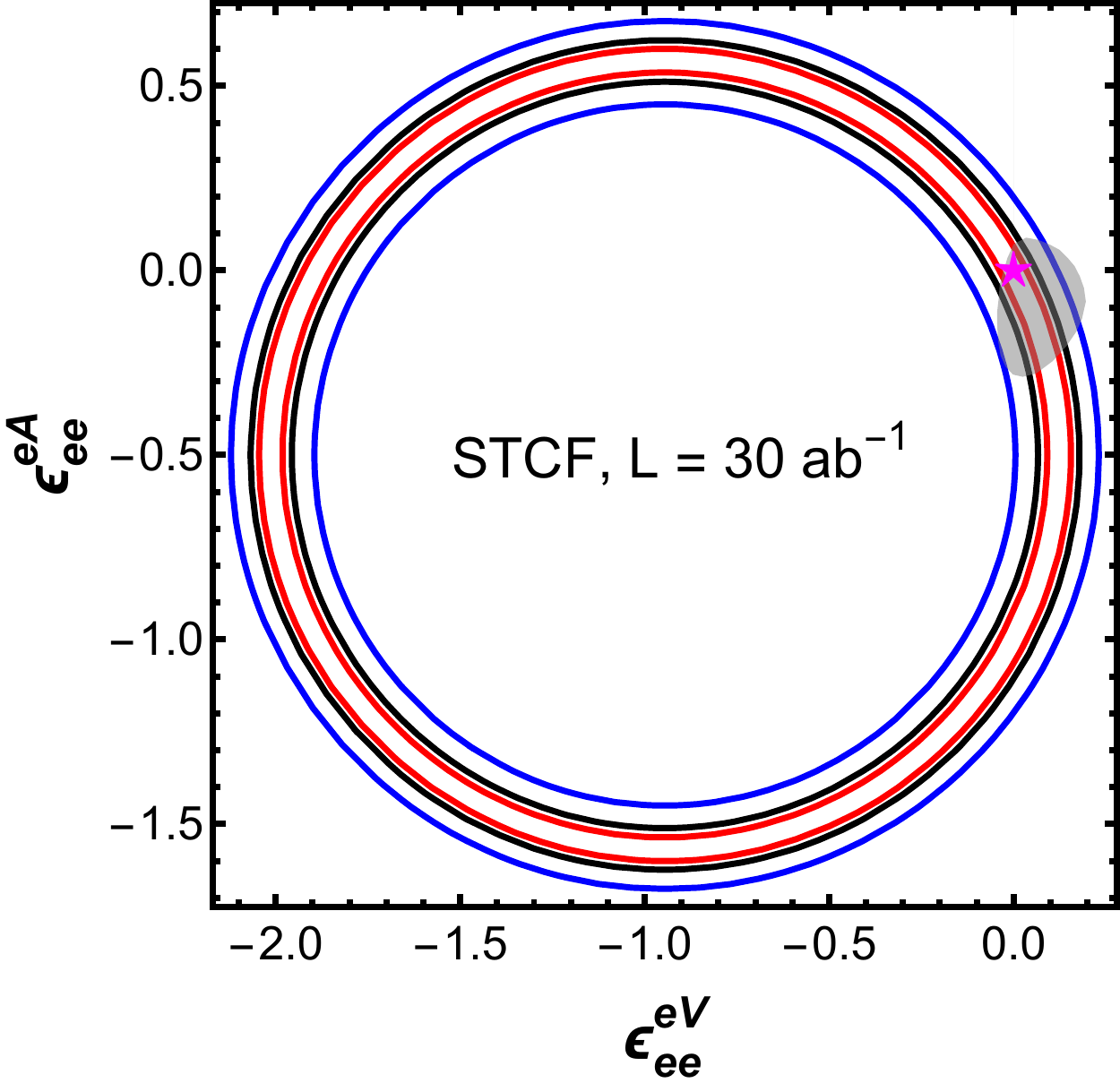}
		\caption{The allowed  90\% C.L. region for electron-type neutrino NSI in the planes of $(\epsilon_{ee}^{eL},\ \epsilon_{ee}^{eR})$ (left panel) and $(\epsilon_{ee}^{eV},\ \epsilon_{ee}^{eA})$ (right panel) at future L = 30 ab$^{-1}$ STCF running with
		$\sqrt{s}=$2 (blue lines), 4 (black lines) and 7 (red lines) GeV, respectively.
	   The  allowed 90\% C.L. regions  arising
	from the global analysis of the LEP, CHARM II, LSND, and reactor data \cite{Barranco:2007ej},  are shown in  the shaded gray regions.}
		\label{fig:stcfee}
	\end{centering}
\end{figure}

\begin{figure}[htbp]
	\begin{centering}
		\includegraphics[width=0.45\columnwidth]{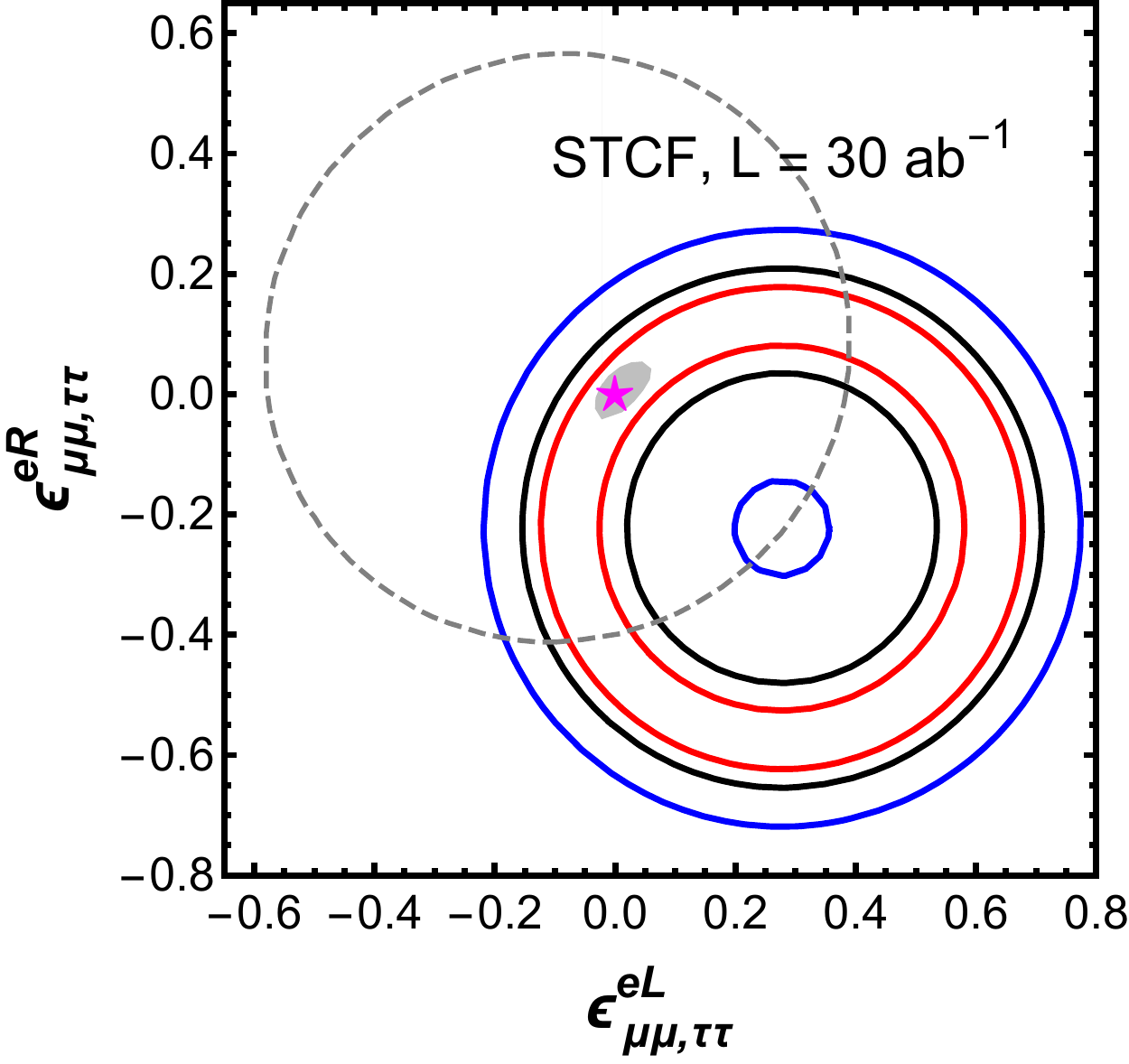}
		\includegraphics[width=0.45\columnwidth]{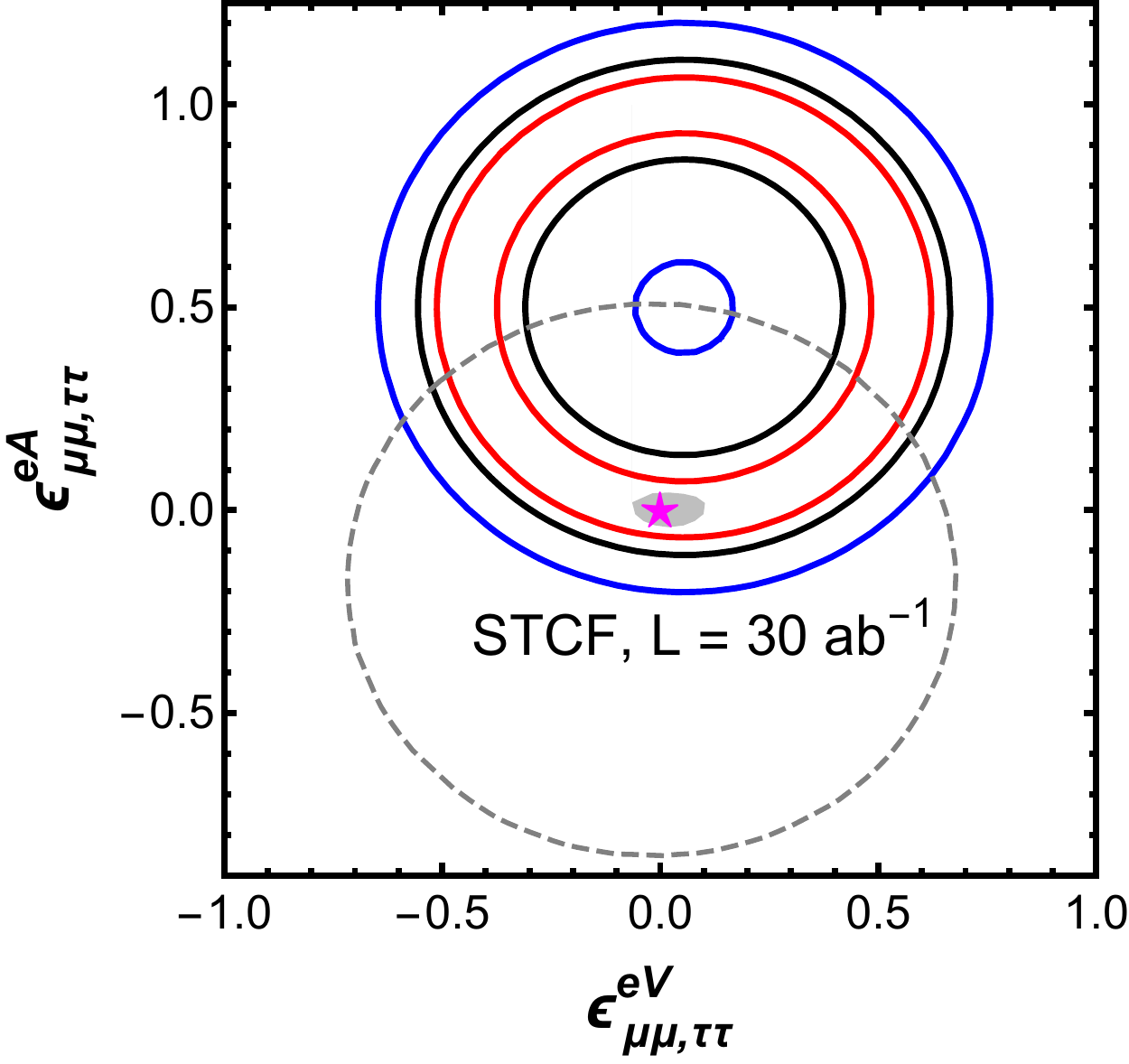}
		\caption{The allowed  90\% C.L. region for muon-type and tau-type neutrino NSI in the planes of $(\epsilon_{\mu\mu,\tau\tau}^{eL},\ \epsilon_{\mu\mu,\tau\tau}^{eR})$ (left panel) and $(\epsilon_{\mu\mu,\tau\tau}^{eV},\ \epsilon_{\mu\mu,\tau\tau}^{eA})$ (right panel) at future L = 30 ab$^{-1}$ STCF running with
			$\sqrt{s}=$2 (blue lines), 4 (black lines) and 7 (red lines) GeV, respectively.
			The  allowed 90\% C.L. regions for muon-type arising
			from the global analysis of the LEP, CHARM II, LSND, and reactor data \cite{Barranco:2007ej},  are shown in the shaded gray regions, and the allowed 95\% C.L. regions for tau-type NSI arising from the LEP data are shown in dashed gray lines.}
		\label{fig:stcfmm}
	\end{centering}
\end{figure}

Using the $\chi^2$ defined in Eq.~(\ref{eq:chi2}) and the relation in Eq.~(\ref{eq:delta}), we can get $\chi^2 = \delta^2 \sigma^{SM}L$  with the ``{\it advanced cuts}" cut
at $L=$ 30 ab$^{-1}$ STCF. We present 90\% C.L. allowed  regions by solving 
$\chi^{2} (\epsilon_{\alpha\alpha}^{eL/V},\ \epsilon_{\alpha\alpha}^{eR/A})-\chi^{2} (0,0)=4.61$ for each specific pair with $\alpha=e,\ \mu,\ \tau$,
in Fig.~\ref{fig:stcfee}  for electron-type and Fig.~\ref{fig:stcfmm} for muon/tau-type NSI parameters, respectively. 
There we consider three typical running energies $\sqrt{s}=2,\ 4,\ 7$ GeV for  STCF.

Allowed region at 90\% C.L. from each STCF experiment is between the two concentric circles.
The centers of the circles are independent of the running energy of STCF, and are also the same as Belle II.
With same integrated luminosity, better constraints can be obtained with higher
running energy at STCF due to the larger cross section for neutrino
produciton. 
These plots clearly indicate 
that future STCF can be complementary with current global analysis
in constraining $(\epsilon_{ee}^{eL},\ \epsilon_{ee}^{eR})$ and $(\epsilon_{ee}^{eV},\ \epsilon_{ee}^{eA})$ and with existing LEP data for tau-type NSI parameter limits, while it will not be competitive for muon-type
NSI parameters.

In Table \ref{tab:stcf}, by varying only one parameter at a time and fixing the remaining parameters to zero,
i.e., by solving 
$\chi^{2} (\epsilon_{\alpha\alpha}^{e})-\chi^{2} (0)=2.71$, we obtain the constraints on each NSI parameter on the neighborhood of the SM point $\epsilon_{\alpha\alpha}^{e}=0$ at $L=$ 30 ab$^{-1}$ STCF with $\sqrt{s}=2,\ 4,\ 7$ GeV and $L=$ 50 ab$^{-1}$ Belle II without considering gBG, respectively.
One can see how future STCF and Belle II experiments
lead to an improvement in the constraints for eletron-type $\epsilon_{ee}^{eL,R}$ and tau-type $\epsilon_{\tau\tau}^{eL,R}$ NSI parameters. The previous constraints on $\epsilon_{ee}^{eL,R}$ coming from global analysis of the LEP, CHARM II, LSND and reactor data, and $\epsilon_{\tau\tau}^{eL,R}$ coming from LEP data can be superseded. However, there is no improvement in the constraints for the parameters $\epsilon_{\mu\mu}^{eL,R}$ at STCF.

\setlength{\tabcolsep}{3mm}{
	\begin{table}
		\renewcommand\arraystretch{1.5}
		\begin{tabular}{c|c|c|c|c|c}
			\hline \hline
			& STCF-2 & STCF-4 & STCF-7 & Belle II &Previous Limit \\ 
			& $L=30$ ab$^{-1}$ & $L=30$ ab$^{-1}$ &  $L=30$ ab$^{-1}$ &  $L=50$ ab$^{-1}$ & 90\% Allowed \cite{Barranco:2007ej} \\ 
			\hline 
			$\epsilon_{ee}^{eL}$ &[-0.067,0.061] &[-0.033,0.031] &[-0.018,0.018] &[-0.0091,0.0089] &[-0.03,0.08] \\
			\hline 
			$\epsilon_{ee}^{eR}$ &[-0.60,0.15] &[-0.163,0.087] &[-0.070,0.053] &[-0.031,0.028] &[0.004,0.15]  \\
			\hline
			$\epsilon_{ee}^{eV}$ &[-0.103,0.093] &[-0.050,0.048] &[-0.028,0.027] &[-0.014,0.014]&  -- \\ 
			\hline 
			$\epsilon_{ee}^{eA}$ &[-0.24,0.16] &[-0.103,0.085] &[-0.056,0.050] &[-0.027,0.026]& -- \\
			\hline 
			$\epsilon_{\mu\mu/\tau\tau}^{eL}$ &[-0.13,0.69] &[-0.073,0.101] &[-0.044,0.052] &[-0.023,0.025] & [-0.03,0.03]/[-0.5,0.2] \\ 
			\hline      
			$\epsilon_{\mu\mu/\tau\tau}^{eR}$ &[-0.60,0.15] &[-0.163,0.087] &[-0.070,0.053] &[-0.031,0.028 ] &[-0.03,0.03]/[-0.3,0.4] \\
			\hline        	
			$\epsilon_{\mu\mu/\tau\tau}^{eV}$ &[-0.38,0.49] &[-0.25,0.36] &[-0.18,0.29] &[-0.12,0.23] & -- \\
			\hline 	
			$\epsilon_{\mu\mu/\tau\tau}^{eA}$ &[-0.16,0.24] &[-0.085,0.103] &[-0.050,0.056] &[-0.026,0.027] & -- \\
			\hline 
			\hline 
		\end{tabular} 
		\caption{Constraints on NSI parameters by varying only one parameter
		at a time at STCF and Belle II with gBG ignored. For comparison, the previous reported results from
	the global analysis with the LEP, CHARM II, LSND, and reactor data are shown in the last column.}
					\label{tab:stcf}
	\end{table}
}

\section{$e^+e^-$ colliders operated with $\sqrt{s}\geq M_Z$}
\label{sec:sgZ}

In this section, we present the sensitivity on the strength of NSI at $\epem$ colliders
operated with $\sqrt{s}\geq M_Z$.
The results with the monophoton searches at LEP have been shown in Refs.~\cite{Berezhiani:2001rs, Barranco:2007ej, Forero:2011zz}.
Here, we will focus on the future projected CEPC~\cite{CEPCStudyGroup:2018ghi}.
Three different running modes have been proposed for CEPC, including 
the Higgs factory mode for the $e^{+} e^{-} \rightarrow Z H$ production with a total luminosity of $\sim 5.6\ \mathrm{ab}^{-1}$ for seven years running
at $\sqrt{s}=240\ \mathrm{GeV}$ , the $Z$ factory mode for the $e^{+} e^{-} \rightarrow Z$ production
with a total luminosity of $\sim 16\ \mathrm{ab}^{-1}$ for two years running at $\sqrt{s}=91.2\ \mathrm{GeV}$ , and the $W^+W^-$ threshold scan mode for the $e^{+} e^{-} \rightarrow W^{+} W^{-}$ production
with a total luminosity of $\sim 2.6\ \mathrm{ab}^{-1}$ for one year running at 
$\sqrt{s} \sim$ $158-172\ \mathrm{GeV}$.
\footnote{We take $\sqrt s= 160$ GeV for the $W^+W^-$ threshold scan mode throughout our analysis.}
For the monophoton signature at CEPC, we use the cuts for the final detected photon 
(hereafter the ``{\it basic cuts}")
following the CEPC CDR~\cite{CEPCStudyGroup:2018ghi}: $|z_\gamma|<0.99$ and $E_\gamma > 0.1 $  GeV. 
Similar as the symmetric STCF, we also apply the cuts (\ref{eq:bBG-besiii}) as  ``{\it advanced cuts}"  
with $\cos\theta_b=0.99$ to remove the reducible background events \cite{Liu:2019ogn}.

\begin{figure}[htbp]
	\begin{centering}
		\includegraphics[width=0.45\columnwidth]{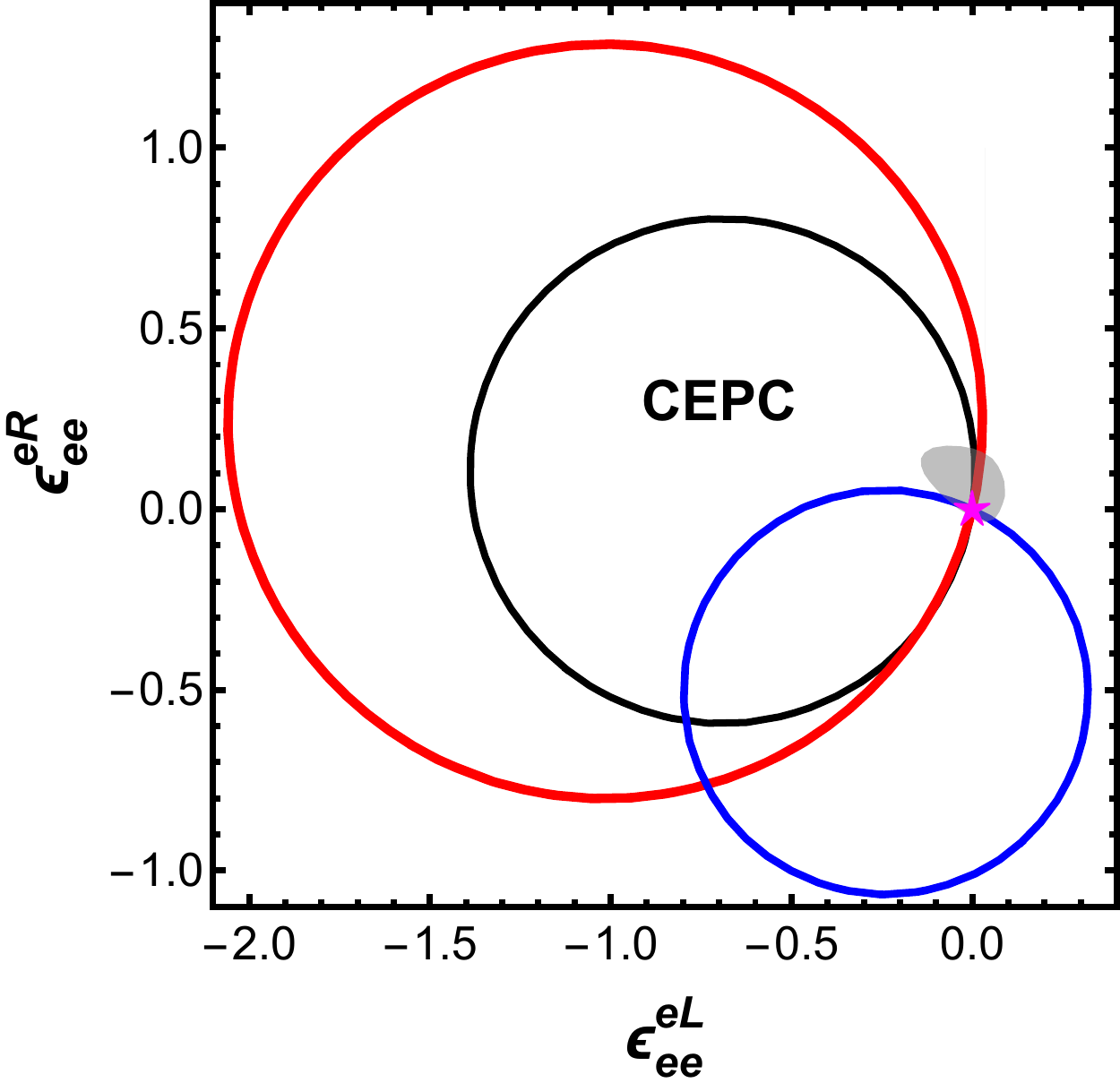}
		\includegraphics[width=0.48\columnwidth]{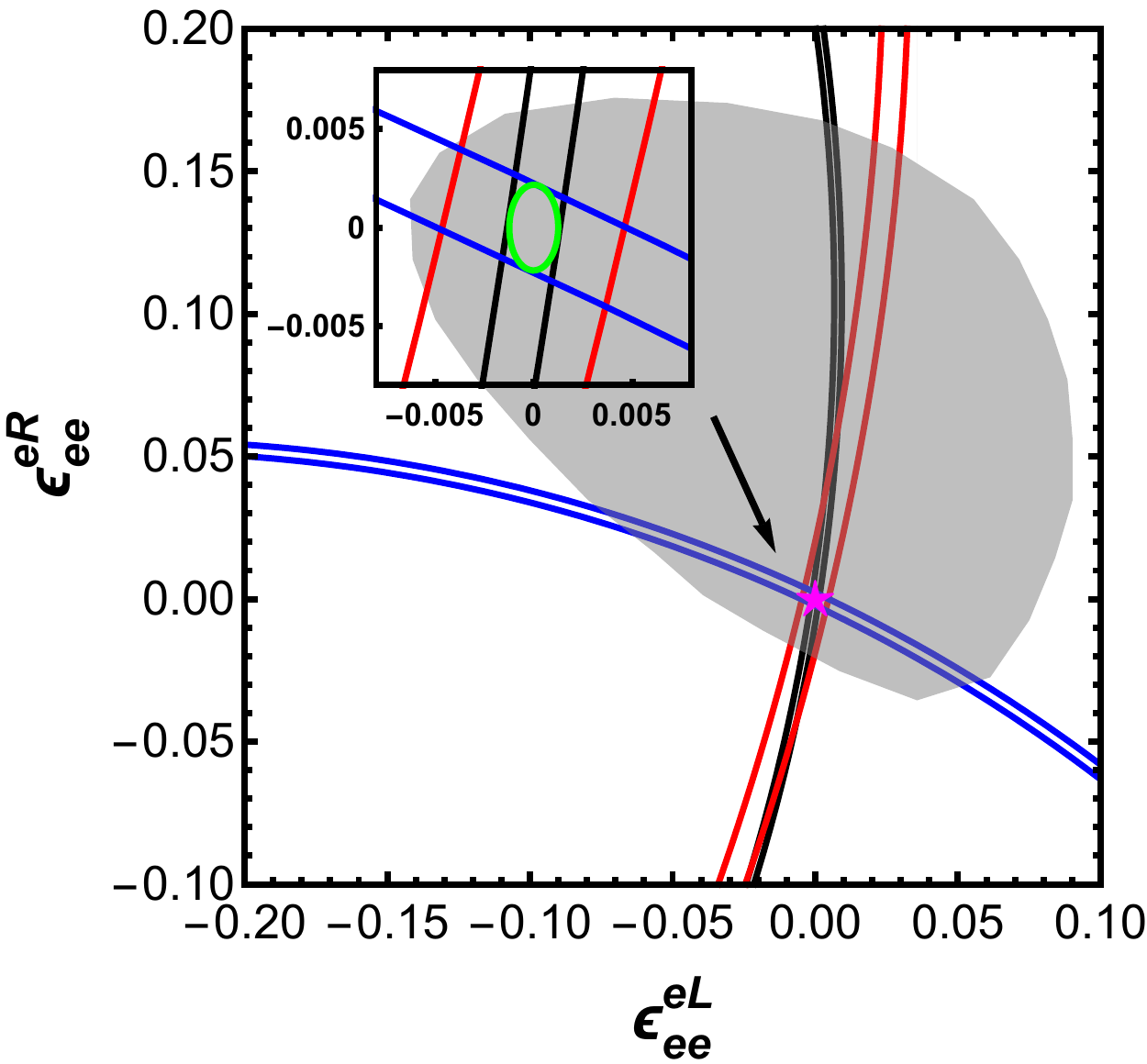}
		\caption{The allowed  90\% C.L. region for electron-type neutrino NSI in the planes of $(\epsilon_{ee}^{eL},\ \epsilon_{ee}^{eR})$ at future CEPC with 5.6 ab$^{-1}$ data of $\sqrt{s}=240$ GeV (Black), with 2.6 ab$^{-1}$ data of $\sqrt{s}=160$ GeV (Red), and with 16 ab$^{-1}$ data of $\sqrt{s}=91.2$ GeV (Blue), respectively.
	   The  allowed 90\% C.L. regions  arising
	from the global analysis of the LEP, CHARM, LSND, and reactor data \cite{Barranco:2007ej},  are shown in the shaded gray regions.
	With all the data collected in all three running modes, the combined result is labelled by green and shown in the right panel.
	}
		\label{fig:cepceelr}
	\end{centering}
\end{figure}

\begin{figure}[htbp]
	\begin{centering}
		\includegraphics[width=0.45\columnwidth]{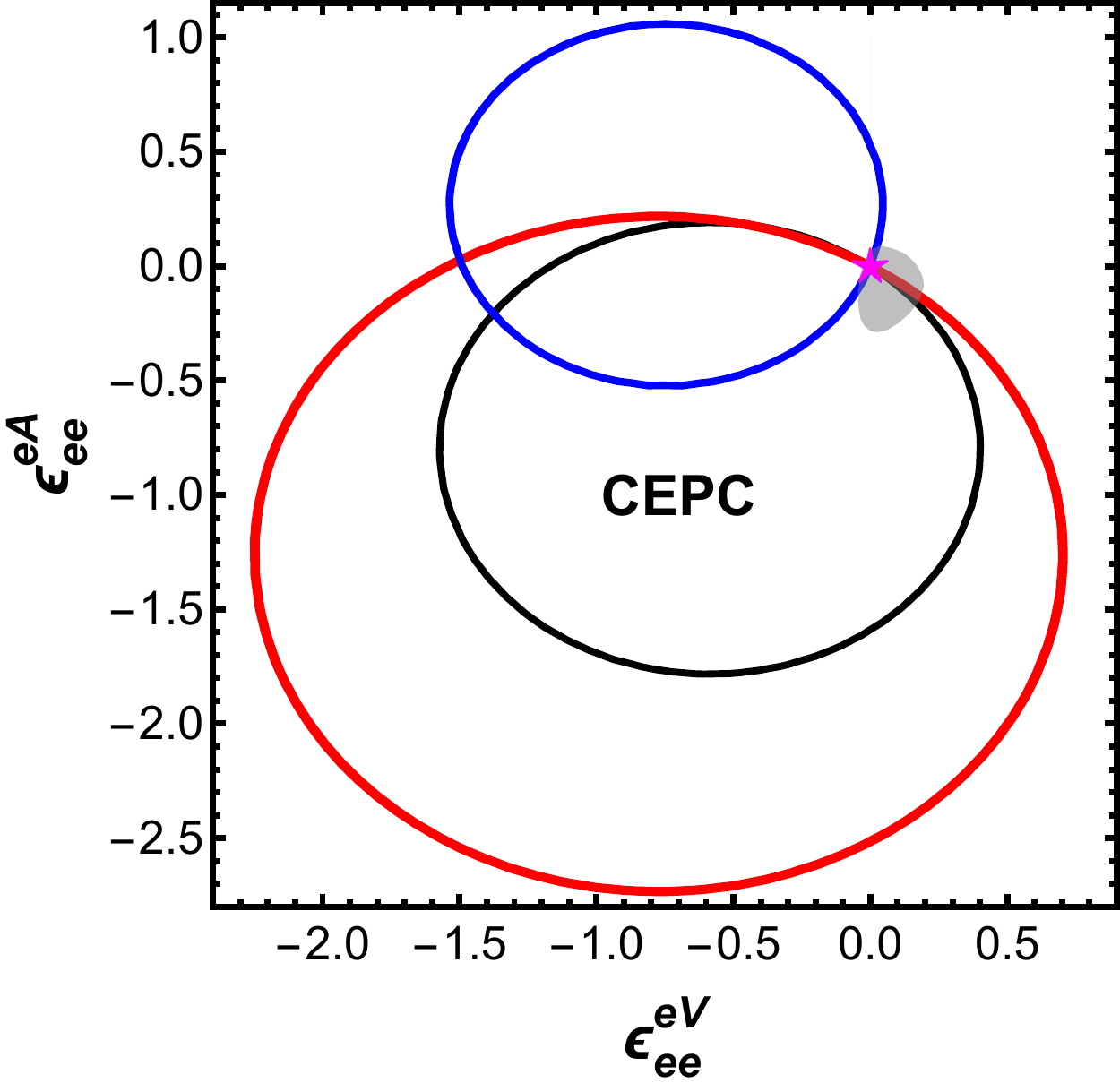}
		\includegraphics[width=0.47\columnwidth]{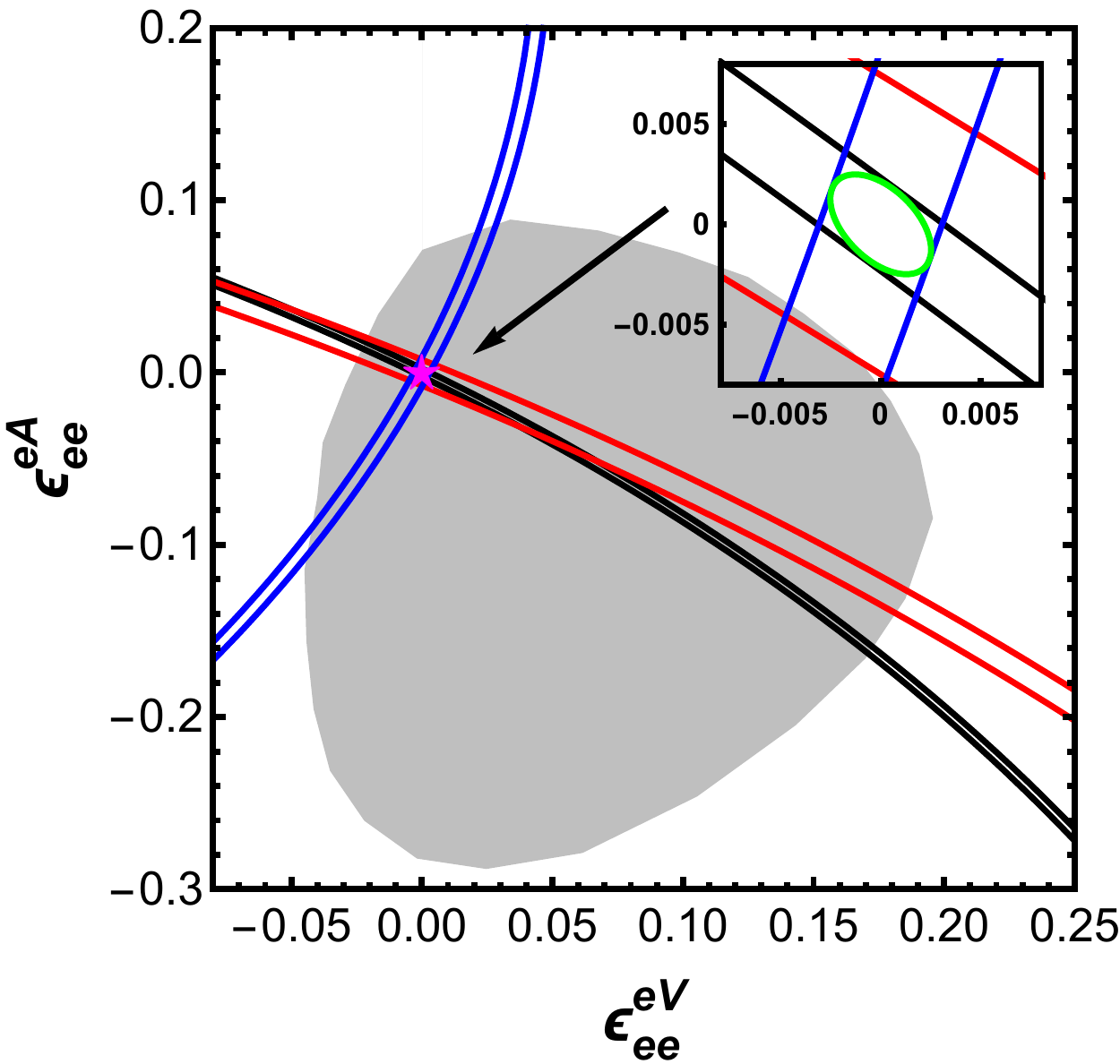}
		\caption{The allowed  90\% C.L. region for electron-type neutrino NSI in the planes of $(\epsilon_{ee}^{eV},\ \epsilon_{ee}^{eA})$ at future CEPC with 5.6 ab$^{-1}$ data of $\sqrt{s}=240$ GeV (Black), with 2.6 ab$^{-1}$ data of $\sqrt{s}=160$ GeV (Red), and with 16 ab$^{-1}$ data of $\sqrt{s}=91.2$ GeV (Blue), respectively.
	   The  allowed 90\% C.L. regions  arising
	from the global analysis of the LEP, CHARM II, LSND, and reactor data \cite{Barranco:2007ej},  are shown in  the shaded gray regions.	With all the data collected in all three running modes, the combined result is labelled by green and shown in the right panel.}
		\label{fig:cepceeva}
	\end{centering}
\end{figure}

\begin{figure}[htbp]
	\begin{centering}
		\includegraphics[width=0.45\columnwidth]{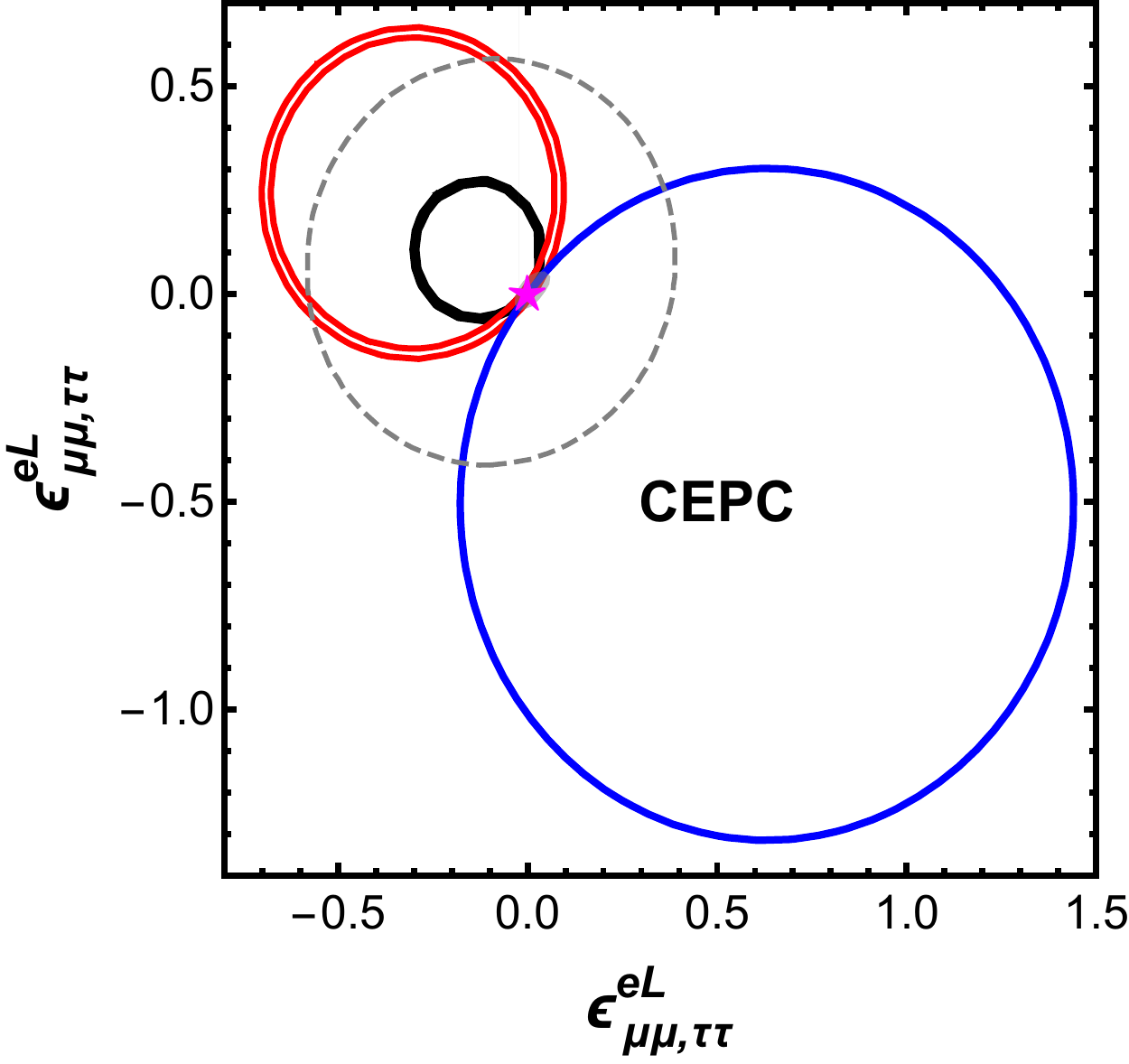}
		\includegraphics[width=0.45\columnwidth]{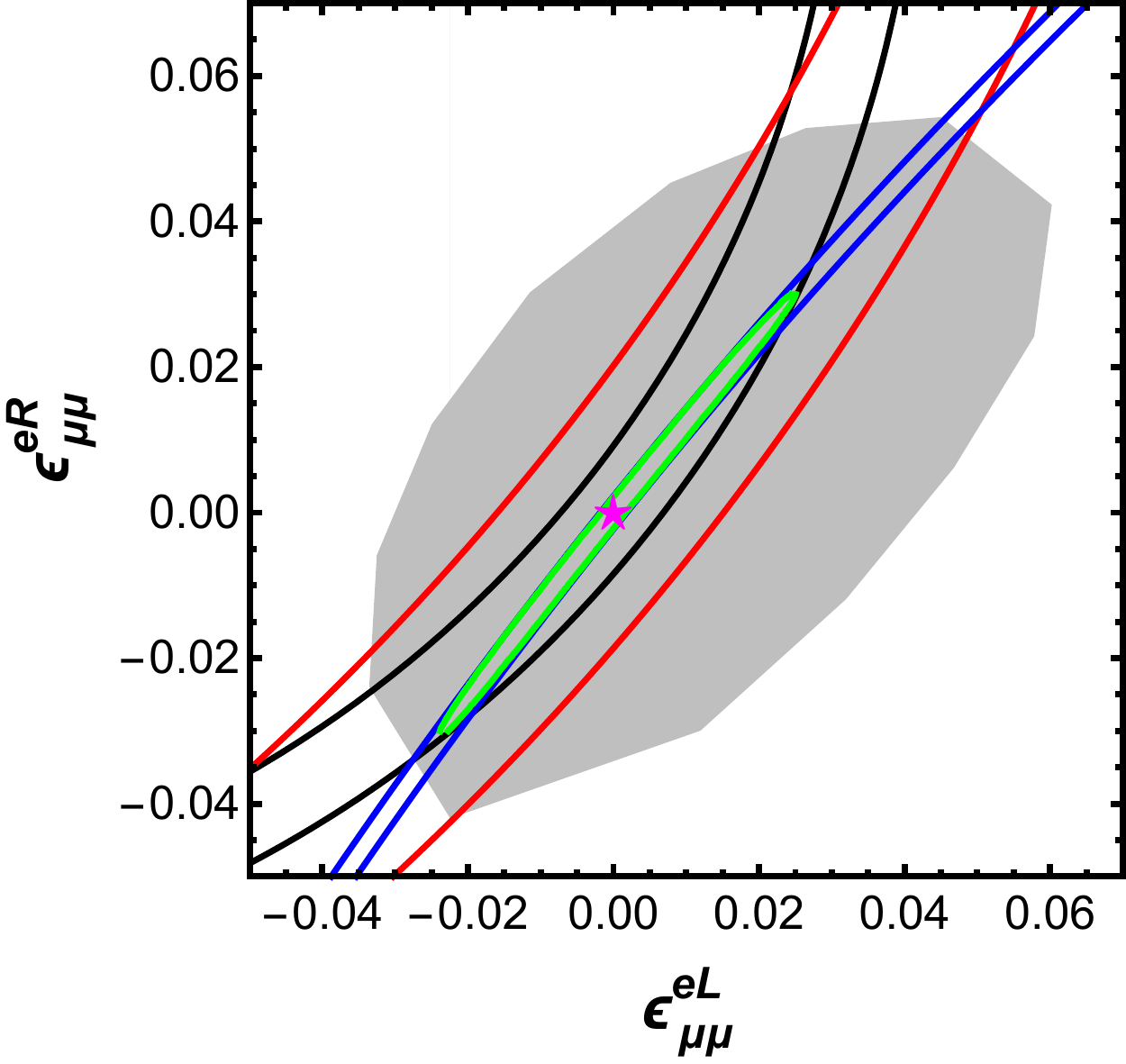}
		\caption{
		The allowed  90\% C.L. region for muon-type and tau-type neutrino NSI in the planes of $(\epsilon_{\mu\mu,\tau\tau}^{eL},\ \epsilon_{\mu\mu,\tau\tau}^{eR})$at future CEPC with 5.6 ab$^{-1}$ data of $\sqrt{s}=240$ GeV (Black), with 2.6 ab$^{-1}$ data of $\sqrt{s}=160$ GeV (Red), and with 16 ab$^{-1}$ data of $\sqrt{s}=91.2$ GeV (Blue), respectively. 
			The  allowed 90\% C.L. regions for muon-type arising
			from the global analysis of the LEP, CHARM II, LSND, and reactor data \cite{Barranco:2007ej},  are shown in the shaded gray regions, and the allowed 90\% C.L. regions for tau-type NSI arising from the LEP data are shown in dashed gray lines. 	With all the data collected in all three running modes, the combined result is labelled by green and shown in the right panel.}
		\label{fig:cepcmmlr}
	\end{centering}
\end{figure}

\begin{figure}[htbp]
	\begin{centering}
		\includegraphics[width=0.45\columnwidth]{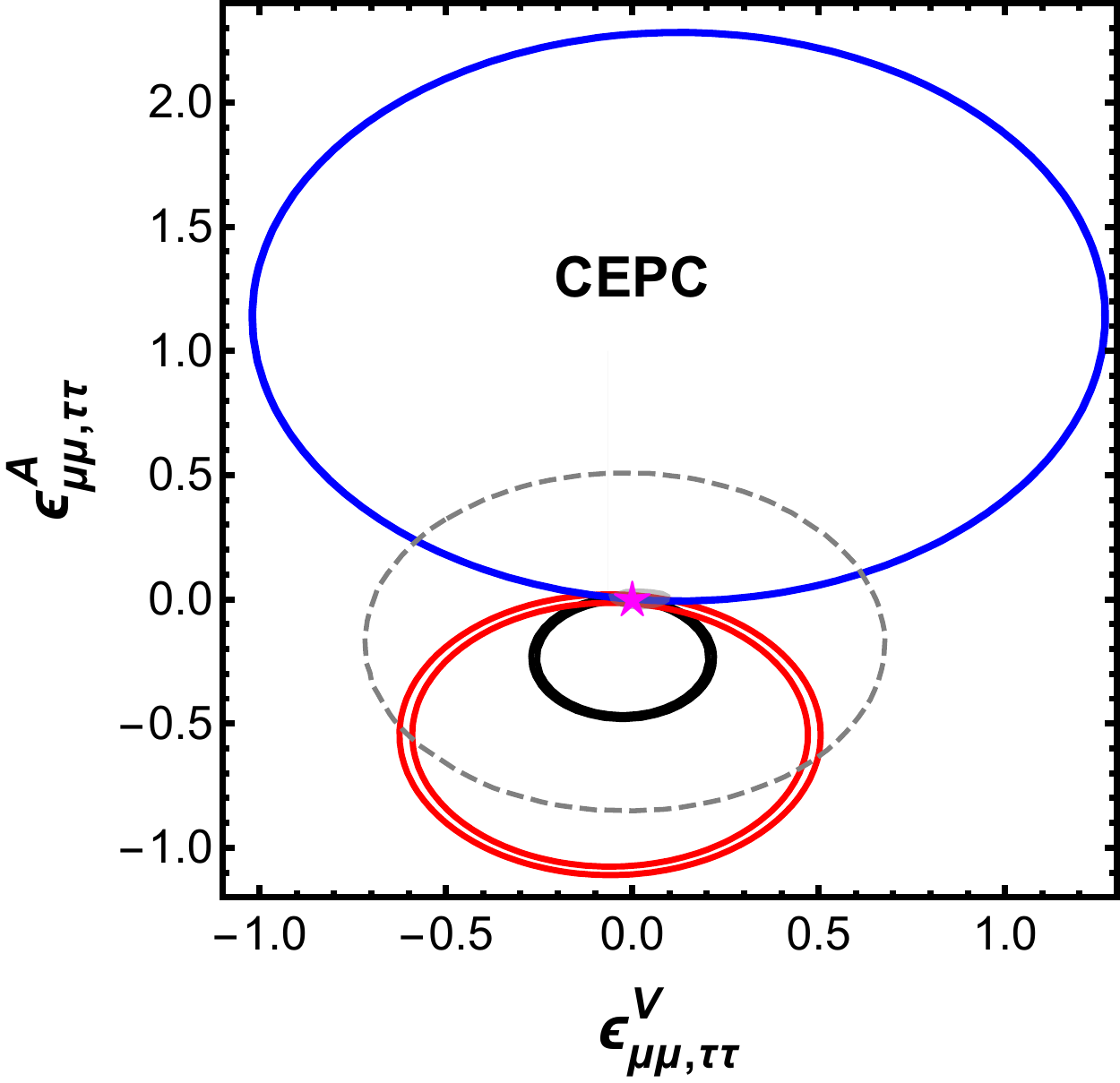}
		\includegraphics[width=0.45\columnwidth]{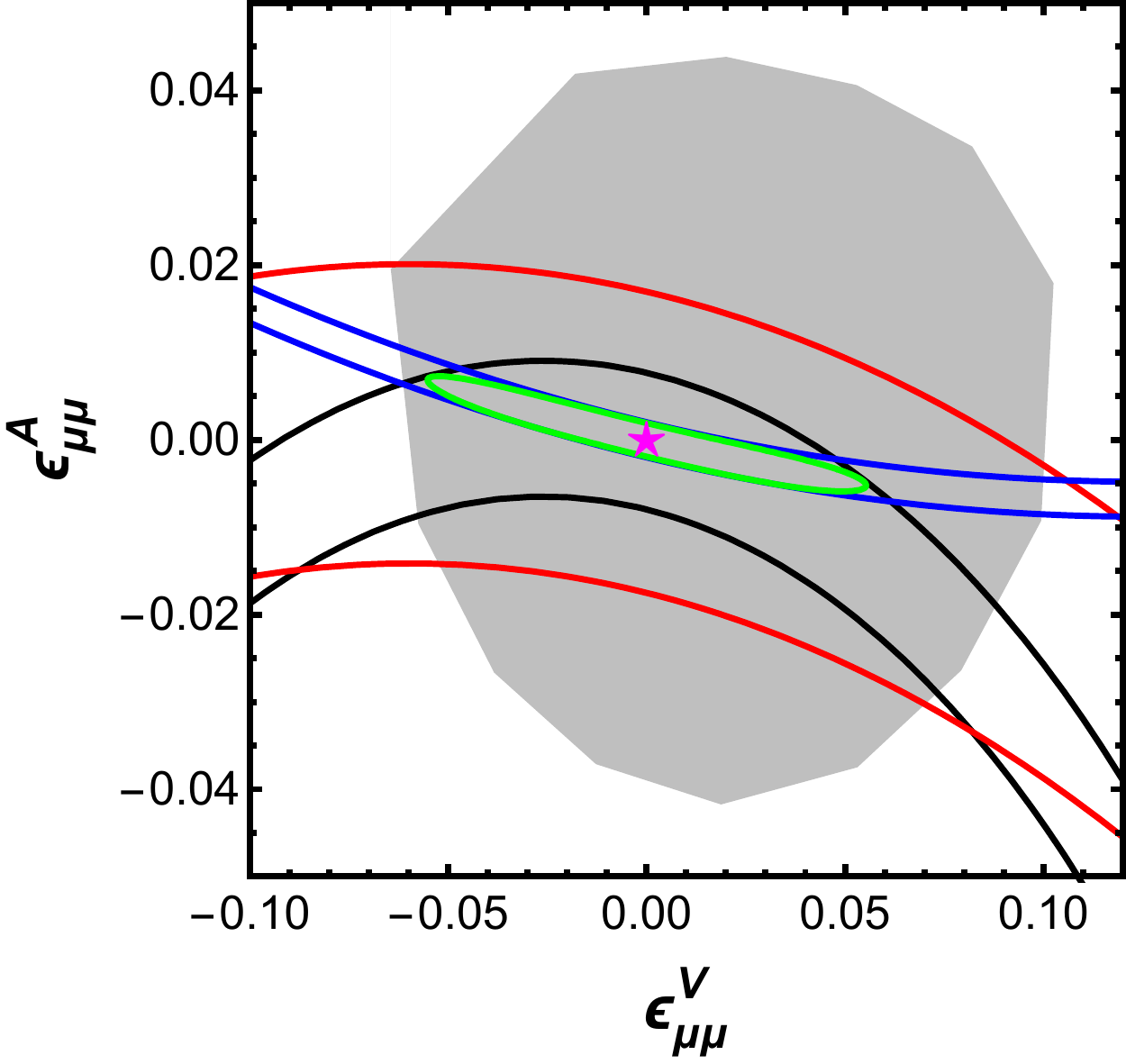}
		\caption{	The allowed  90\% C.L. region for muon-type and tau-type neutrino NSI in the planes of $(\epsilon_{\mu\mu,\tau\tau}^{eV},\ \epsilon_{\mu\mu,\tau\tau}^{eA})$at future CEPC with 5.6 ab$^{-1}$ data of $\sqrt{s}=240$ GeV (Black), with 2.6 ab$^{-1}$ data of $\sqrt{s}=160$ GeV (Red), and with 16 ab$^{-1}$ data of $\sqrt{s}=91.2$ GeV (Blue), respectively. 
			The  allowed 90\% C.L. regions for muon-type arising
			from the global analysis of the LEP, CHARM II, LSND, and reactor data \cite{Barranco:2007ej},  are shown in the shaded gray regions, and the allowed 90\% C.L. regions for tau-type arising from the LEP data are shown in dashed gray lines. 	With all the data collected in all three running modes, the combined result is labelled by green and shown in the right panel.}
		\label{fig:cepcmmva}
	\end{centering}
\end{figure}

Integrating out the final photon with the ``{\it advanced cuts}" for the differential cross section of Eq.~(\ref{eq:diffxs}) with 
the radiator function and the neutrino pair production cross section from SM of Eq. (\ref{eq:xs0sm}) or from 
NSI of Eq.~(\ref{eq:xs0nsi}), 
we can get the corresponding total cross section. 
Using the $\chi^2$ defined in Eq. (\ref{eq:chi2}), and solving 
$\chi^{2} (\epsilon_{\alpha\alpha}^{eL/V},\ \epsilon_{\alpha\alpha}^{eR/A})-\chi^{2} (0,0)=4.61$,
the 90\% C.L. allowed regions for each specific pair with $\alpha=e,\ \mu,\ \tau$
are shown in Figs. (\ref{fig:cepceelr}-\ref{fig:cepcmmva}), 
which lie between the two concentric circles lablled black for $\sqrt{s}=240\ \mathrm{GeV}$,
red for  $\sqrt{s}=160\ \mathrm{GeV}$ and blue for $\sqrt{s}=91.2\ \mathrm{GeV}$, respectively.
We also show the allowed regions by combining the data from the three different running modes with planned luminosity, which are enclosed by the blue curves in right panels of Figs.~(\ref{fig:cepceelr}-\ref{fig:cepcmmva}). As we see from Figs.~(\ref{fig:cepceelr}-\ref{fig:cepcmmva}), after combining the data from three different modes, the allowed regions for all NSI parameters can be severely constrained as compared to the global analysis. 

\begin{figure}[htbp]
	\begin{centering}
		\includegraphics[width=0.45\columnwidth]{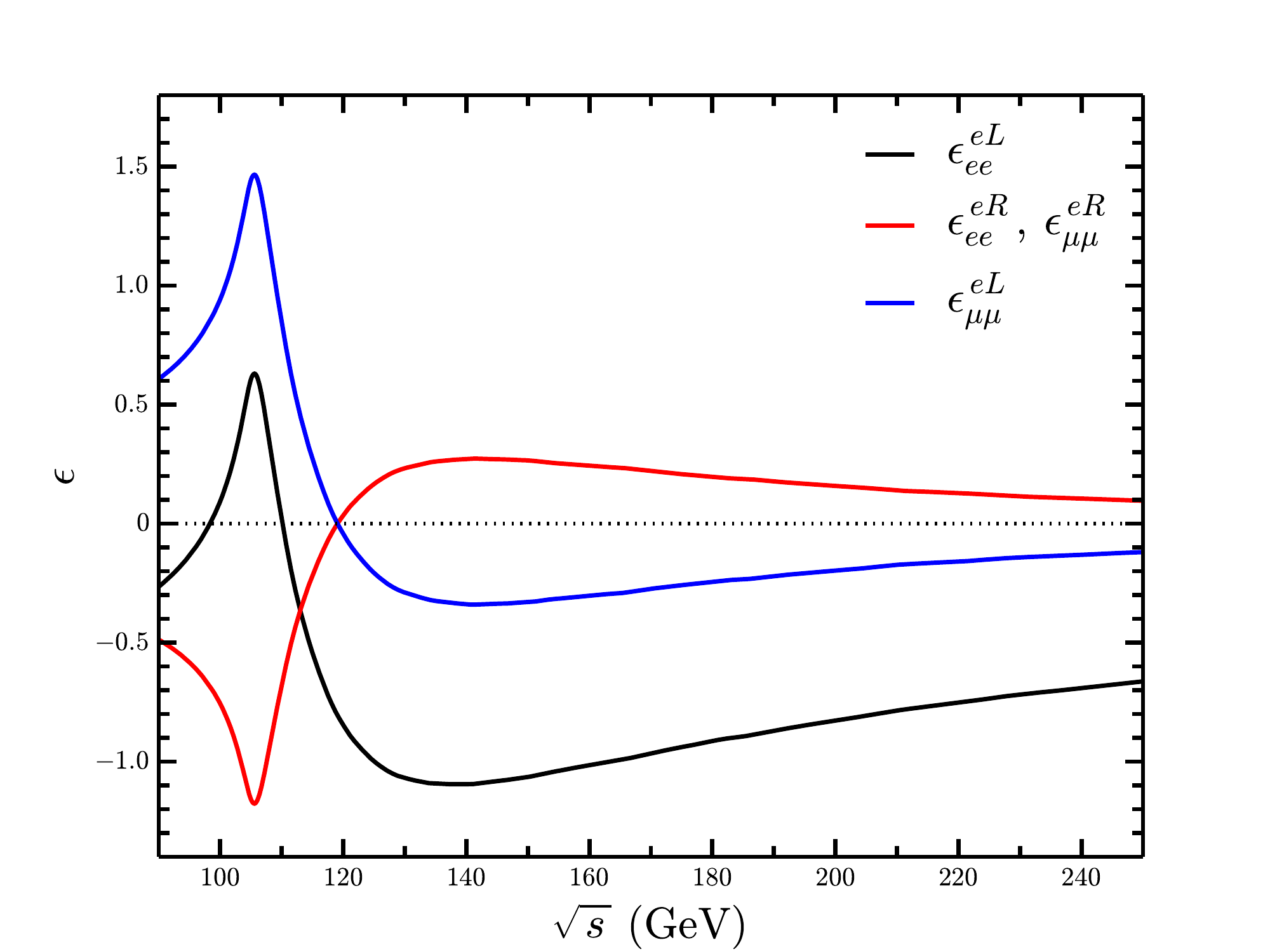}
		\includegraphics[width=0.45\columnwidth]{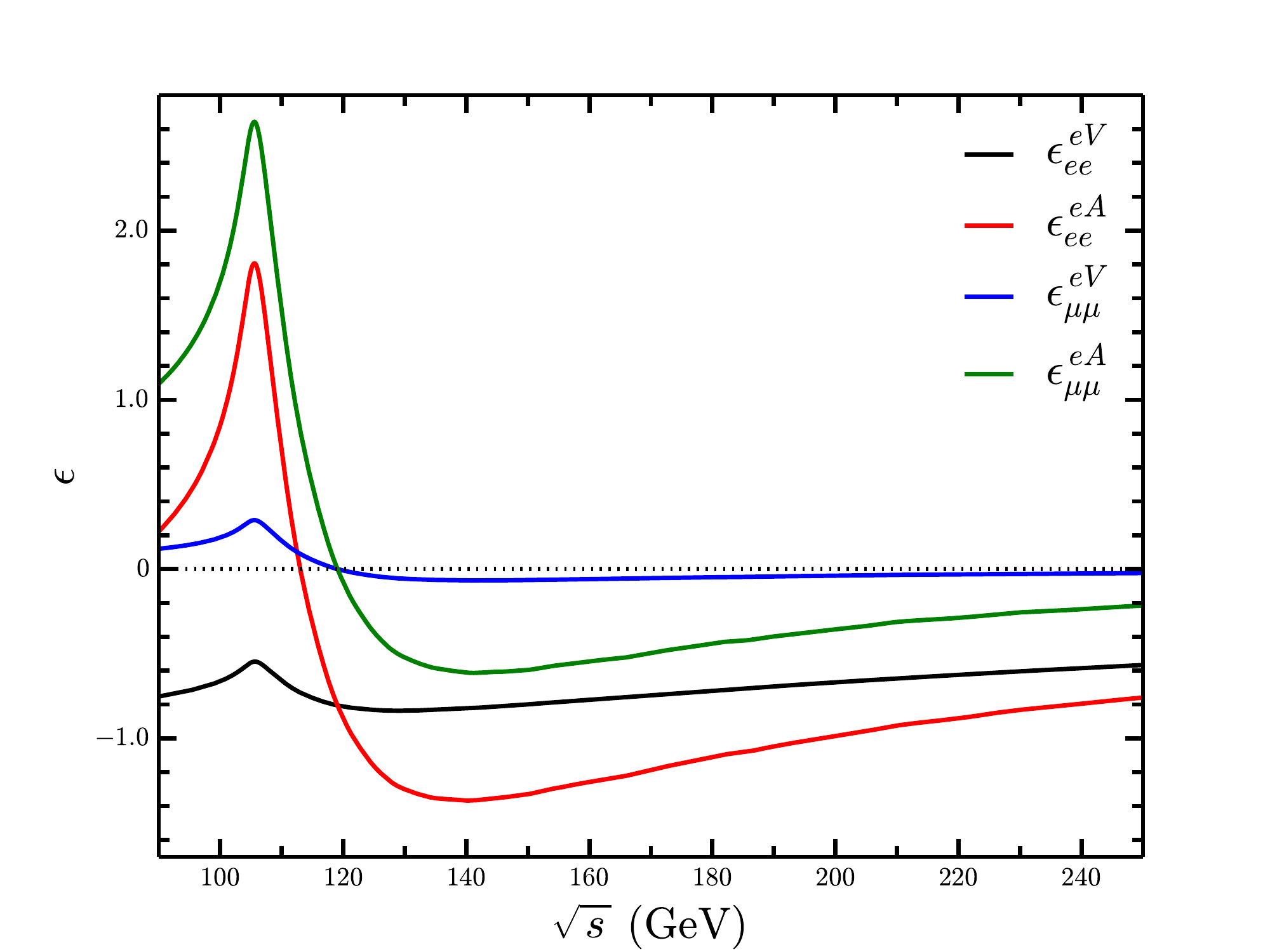}
		\caption{The coordinates of the circle centers for each pair of NSI parameter $(\epsilon_{\alpha\alpha}^{eL},\ \epsilon_{\alpha\alpha}^{eR})$ (Left) and $(\epsilon_{\alpha\alpha}^{eV},\ \epsilon_{\alpha\alpha}^{eA})$ (Right) as a function of $\sqrt{s}$ at CEPC with the ``{\it advanced cuts}".}
		\label{fig:center}
	\end{centering}
\end{figure}

These strong constraints on the NSI parameter space can be understood as follows.
Unlike the STCF and Belle II with $\sqrt{s} \ll M_Z$, the centers of the iso-$\sigma$ circles with different running energy no longer stay the same but become a function of $\sqrt{s}$. 
By integrating out the final photon with the ``{\it advanced cuts} from Eqs. (\ref{eq:center-e-high}) and (\ref{eq:center-mu-high}),
 we can get the coordinates of the circle center for each pair of NSI parameter $(\epsilon_{\alpha\alpha}^{eL},\ \epsilon_{\alpha\alpha}^{eR})$ or $(\epsilon_{\alpha\alpha}^{eA},\ \epsilon_{\alpha\alpha}^{eV})$ as the function of $\sqrt{s}$, which are shown in Fig. \ref{fig:center}.

Since we simulate the data assuming the SM, the point (0,0) will always be included in the allowed regions. Also, the allowed regions from each running mode will lie between two concentric circles. From Fig.~(\ref{fig:cepceelr}-\ref{fig:cepcmmva}), we see that the direction from the SM point (0,0) to the circle center with $\sqrt{s}=240\ \mathrm{GeV}$ is almost parallel to that with $\sqrt{s}=160\ \mathrm{GeV}$, and the allowed regions with $\sqrt{s}=160\ \mathrm{GeV}$ can be superseded by $\sqrt{s}=240\ \mathrm{GeV}$ in the neighborhood of the SM point (0,0). For electron-type NSI, 
the direction from the SM point (0,0) to the circle center with $\sqrt{s}=91.2\ \mathrm{GeV}$
is approximately perpendicular to that with the other two running mode. Therefore, the allowed parameter space of electron-type NSI are severely constrained by combining the data from three running modes. Even if both $\epsilon_{ee}^{eL}$ and $\epsilon_{ee}^{eR}$ are present, the allowed ranges for $|\epsilon_{ee}^{eL}|$ or $|\epsilon_{ee}^{eR}|$ can be constrained to be smaller than 0.002. 
As for muon-type or tau-type NSI, the direction from the SM point (0,0) to the circle center with $\sqrt{s}=91.2\ \mathrm{GeV}$ is approximately opposite to that with the other two running modes. Hence, the allowed regions will become long ellipses after combining the data from three running modes. If both $\epsilon_{\mu\mu,\tau\tau}^{eL}$ and $\epsilon_{\mu\mu,\tau\tau}^{eR}$ are present, the allowed ranges of muon-type and tau-type NSI are about one order of magnitude weaker than that of electron-type NSI.


In Table \ref{tab:cepc}, we present the 90\% C.L. constraints on only one parameter at a time by fixing the remaining parameters to zero at CEPC. One can find that with three different running modes, CEPC can
lead to a great improvement in constraining each NSI parameter compared to the previous global analysis of the LEP, CHARM II, LSND and reactor data~\cite{Barranco:2007ej}.
Except for $\epsilon_{\mu\mu,\tau\tau}^{eV}$, one can constrain all NSI parameters to be smaller than 0.002 by combining the data from three running modes. 


\setlength{\tabcolsep}{2mm}{
	\begin{table}
		\renewcommand\arraystretch{1.5}
		\begin{tabular}{c|c|c|c|c|c}
			\hline \hline
			& CEPC-91.2 & CEPC-160 & CEPC-240 & CEPC-combined & Previous Limit\\ 
			&  $L=16$ ab$^{-1}$ &  $L=2.6$ ab$^{-1}$ & $L=5.6$ ab$^{-1}$ & $L=24.2$ ab$^{-1}$ & 90\% Allowed  \cite{Barranco:2007ej} \\ 
			\hline 
			$\epsilon_{ee}^{eL}$  &[-0.0037,0.0037]&[-0.0036,0.0035] &[-0.0010,0.0010] & [-0.00095,0.00095] &[-0.03,0.08]\\
			\hline 
			$\epsilon_{ee}^{eR}$  &[-0.0017,0.0017] &[-0.014,0.015] &[-0.0065,0.0070] &  [-0.0017,0.0017] & [0.004,0.15]\\
			\hline
			$\epsilon_{ee}^{eV}$   &[-0.0024,0.0023] &[-0.0094,0.0093] &[-0.0024,0.0024]  & [-0.0017,0.0017] & --\\ 
			\hline 
			$\epsilon_{ee}^{eA}$  &[-0.0065,0.0066] &[-0.0057,0.0057] &[-0.0018,0.0018]  & [-0.0017,0.0017] & --\\
			\hline 
			$\epsilon_{\mu\mu/\tau\tau}^{eL}$  &[-0.0014,0.0014] &[-0.012,0.012] &[-0.0055,0.0053] &[-0.0013,0.0013] &[-0.03,0.03]/[-0.5,0.3] \\
			\hline      
			$\epsilon_{\mu\mu/\tau\tau}^{eR}$  &[-0.0017,0.0017] &[-0.014,0.015] &[-0.0065,0.0070] &[-0.0017,0.0017] &[-0.03,0.03]/[-0.3,0.4]\\
			\hline        	
 			$\epsilon_{\mu\mu/\tau\tau}^{eV}$  &[-0.013,0.015] &[-0.194,0.074] &[-0.085,0.033]   & [-0.013,0.014] & --\\
 			\hline 	
 			$\epsilon_{\mu\mu/\tau\tau}^{eA}$  &[-0.0015,0.0015] &[-0.013,0.013] &[-0.0061,0.0060] &[-0.0015,0.0015] & --\\
			\hline 	 \hline 
		\end{tabular} 
		\caption{Constraints of NSI parameters by varing only one parameter
		at a time at CEPC. For comparasion, the previous reported results from
	the global analysis with the data of the LEP, CHARM II, LSND, and reactor pieces are shown in the last coloumn.}
		\label{tab:cepc}
	\end{table}
}

\section{Conclusions}
\label{sec:Con}

Neutrino oscillation experiments will be affected by the presence of NSI, and it is natural to seek complementary constraints on NSI from other experiments. In this work, we investigate the constraints on NC NSI with electrons at current $\epem$ collider experiment Belle II, and the proposed future $\epem$ collider experiments STCF and CEPC. 
The allowed regions in the $(\epsilon_{\alpha\alpha}^{eL},\ \epsilon_{\alpha\alpha}^{eR})$ (or $(\epsilon_{\alpha\alpha}^{eV},\ \epsilon_{\alpha\alpha}^{eA})$) plane from these $\epem$ collider experiments are shown in Figs.~(\ref{fig:belle2ee}-\ref{fig:cepcmmva}).
The constraints on one NSI parameter at a time are listed in Table.~\ref{tab:stcf} and~\ref{tab:cepc}.

For $\epem$ collider experiments with CM energy $\sqrt{s}\ll M_Z$, we find that the centers of the iso-$\sigma$ circles are independent of the operating energy. Both Belle II and STCF can provide competitive and complementary bounds on electron-type NSI parameters as compared to current global analysis, and strong improvements in the constraints on tau-type NSI. Belle II with gBG ignored will also yield better constraints than STCF due to the larger cross section for neutrino production. However, due to the cancellations between the left-handed (vector) and right-handed (axial-vector) NSI parameters, the allowed ranges on the NSI parameters from Belle II and STCF will be still large if multiple NSI parameters are present. 

For $\epem$ collider experiments with $\sqrt{s}\geq M_Z$, the centers of the iso-$\sigma$ circles are dependent on $\sqrt{s}$. We find that future CEPC experiment with three CM energies $\sqrt{s}=240, 160, 91.2$ GeV will break the degeneracy between the left-handed (vector) and right-handed (axial-vector) NSI parameters after combining the data from the three running modes, and the allowed parameter space of NSI with electrons will be severely constrained even if both the left and right-handed NSI parameters are present. By combining the data from the three running modes, we find that the allowed ranges for $|\epsilon_{ee}^{eL}|$ or $|\epsilon_{ee}^{eR}|$ can be smaller than 0.002, and the allowed ranges for muon-type and tau-type NSI are about one order of magnitude weaker than electron-type NSI if both the left and right-handed NSI parameters are present.

\acknowledgments
JL is supported by the National Natural Science Foundation of China under Grant No. 11905299, Guangdong Basic and Applied Basic Research Foundation under Grant No. 2020A1515011479, the  Fundamental  Research  Funds  for  the  Central Universities, and the Sun Yat-Sen University Science Foundation.
YZ is supported by National Science Foundation of China under Grant No. 11805001 and No. 11935001.

\bibliography{refs}

\end{document}